\documentstyle[preprint,aps]{revtex}

\begin{document}

\draft

\title{Theory of imaging a photonic crystal with transmission near-field 
optical microscopy}

 \author{Garnett W. Bryant, Eric L. Shirley and Lori S. Goldner
}

\address{\it National Institute of Standards and Technology,
Gaithersburg, MD 20899\\
}

 \author{Eric B. McDaniel and J. W. P. Hsu
}

\address{\it Department of Physics, University of Virginia, Charlottesville, 
Virginia 22901 \\
}

 \author{R. J. Tonucci
}

\address{\it Naval Research Laboratory, Washington, DC 20375 \\
}
\date{Current version dated \today }
\maketitle

 \begin{abstract} 
 \noindent While near-field scanning optical microscopy (NSOM) 
can provide optical images with resolution much better
 than the diffraction limit, analysis and interpretation
of these images is often difficult. 
We present a theory of imaging with transmission 
NSOM that includes the effects of tip field, tip/sample coupling, 
light propagation through the sample and light collection. We apply 
this theory to analyze experimental NSOM images
 of a nanochannel glass (NCG) array obtained in transmission mode. The NCG is
 a triangular array of dielectric rods in a dielectric glass matrix 
with a two-dimensional photonic band structure. We determine the 
modes for the NCG photonic crystal and simulate the
observed data. The calculations show large contrast 
at low numerical aperture ($NA$) of the collection optics and 
detailed structure at high $NA$ consistent
with the observed images. We present calculations as a function of 
$NA$ to identify how the NCG photonic modes contribute to and determine 
the spatial structure in these images.
Calculations are presented as a function of tip/sample position, 
sample index contrast and geometry, and aperture size to identify 
the factors that determine image formation with
transmission NSOM in this experiment.

 \end{abstract}
 

%\begin{multicols}{2}
%\narrowtext

\section{Introduction}

Near-field scanning optical microscopy (NSOM) is an exciting new
class of optical microscopies which can provide optical
resolution much better than the diffraction limit 
\cite{gr1,gr2,gr3,gr4,gr5,gr6}. One 
realization of NSOM uses an aperture that is much smaller
than the wavelength, $\lambda$, of the light as a 
nearly point-like (on the scale of $\lambda$) light source. 
Typically an optical fiber is pulled to a 20-100 nm tip and metal coated,
leaving a small hole in the metal coating at the tip to 
provide a nanometer-scale aperture.
This aperture is then placed very close to the sample surface so 
that light emitted from the aperture does not diffract significantly 
before reaching the sample and superresolution, well below the diffraction 
limit $\lambda/2$, can be
achieved. Components of the light that are strongly localized laterally
by the aperture are evanescent, decaying rapidly as they move 
away from the aperture. Tip/sample distance can be adjusted to control
the contribution from these evanescent tip fields.

NSOM images can now be readily obtained. A key step to the further development 
and application of NSOM is learning how to interpret, understand, and 
analyze these images. In NSOM, the excitation, the detection, 
or both can occur in the near field.
Strong coupling between the sample and the light source/detector, 
which is not present in far-field optical microscopy, will occur 
in NSOM. For example, in transmission (reflection) NSOM, light from 
the metallized fiber tip couples to the sample in the near field of the tip, 
while the light transmitted (reflected) from the sample is collected in 
the far field. The NSOM images are influenced both by the strong 
sample/source coupling and by the far-field optics of the 
collection process. To understand these images one must know how the
localized source field is influenced by the presence of the sample in
the near field, how light scatters from the sample, and how light is 
collected in the far field.
In this paper, we present a theory for imaging with transmission near-field
optical microscopy. Our goal was to develop a theory capable of describing
all parts of the imaging process. A complete theory is necessary to clearly
identify and separate the contributions made by each step in the imaging
process. 

We apply this theory to understand transmission NSOM images 
\cite {hsu} made recently of a nanochannel-glass (NCG) array \cite {ton}.
These images probe the optical mode structure of this two-dimensional (2D) 
photonic crystal \cite {pbs}. The NCG array studied is a
2D triangular array of glass rods in a matrix made from a lower-index 
glass. By heating and pulling this structure the lattice
constants of the array can be controlled. High quality NCGs
with lattice spacings on the order of or much less than 
$\lambda$ can be made. The surfaces that are scanned have nearly flat
topography. For these reasons, NCGs provide
excellent samples for testing NSOM. 
A comparison between theory and experiment
will identify which features of NSOM are important in this experiment,  
what determines the observed contrast and resolution in the transmission 
NSOM images, and which features in the images depend on the details of the tip 
field. 

The index variation of the NCG array is periodic in the 2D plane 
perpendicular to the glass rods, so the array possesses the optical modes
of a 2D photonic crystal. Because the glass rods have the higher index, 
the distribution of these photonic modes in this 2D plane 
tends to be larger in the glass rods. However the index mismatch between 
core and matrix glasses is small for the sample studied, so significant 
coupling of these photonic modes among neighboring cores occurs as well. 
A comparison between theory 
and experiment will show what information about the 2D spatial distribution 
of the NCG photonic crystal modes can be learned  
from imaging with transmission NSOM. 

We briefly describe in Sec. II the experiment 
\cite {hsu} to be modeled and the key findings that 
we wish to understand. The theory used to model images obtained by 
transmission NSOM
is presented in Sec. III. Each step in the process is modeled, including the
tip fields, the optical modes of the sample, field propagation from the source 
through the sample to the detector, and collection of the transmitted light
by the detector. The model used for the nanochannel glass array is also
described in this section. In Sec. IV the calculations done
to simulate the experiments are presented. The results show that the theory
provides a good description of the observed images. The results allow us to
determine which features of the near-field optics most affect
these experiments.
In this section we also present simulations for transmission NSOM of samples 
with much higher index contrast than that of the samples studied 
experimentally. Stronger coupling of the near field to the sample photonic
modes is achieved by increasing the index contrast in the NCG.  
For this case we clearly identify those features in the transmission 
NSOM that arise from strong coupling of the source to the sample in the 
near field and determine how to enhance this coupling. Conclusions are 
presented in Sec. V.

\section{Experimental results}

The transmission NSOM experiment we model is shown schematically in 
Fig.~\ref{f1}. A metal coated fiber NSOM tip was placed about 10 nm
from the sample surface. Light transmitted through the sample was 
collected in the far field by an objective with numerical aperture 
$NA$ \cite{na}.  The nanochannel glass array sample was scanned in $x$ 
and $y$ at constant
separation above the tip to produce a 2D image of the NCG. The images were 
taken with two different wavelengths of light ($\lambda = 670$ nm and
 $\lambda = 488$ nm). Polarization of the light leaving the
 fiber tip was controlled by use of fiber paddles. Light was collected 
for three different $NA$ ($NA = 0.28, 0.55, 0.7$). 

The nanochannel glass sample studied in the experiment \cite {hsu} 
was a 2D triangular 
lattice of one glass (channel glass) embedded in another glass (matrix 
glass) with similar 
but slightly lower index of refraction. The channel glass was cylindrical,
approximately 745 nm in diameter with center-to-center nearest neighbor
 separation of $1.07\pm0.05$ $\mu$m. The index of refraction of the matrix
glass was 1.65-1.68 and that of the channel glass was 0.2\% to 1.2\% higher in the
visible range. 

The detailed results for this experiment, including the gray-scale images 
made by transmission NSOM and line scans from these gray-scale images, are
presented in Ref. \cite {hsu}. We summarize here the key findings. For low
$NA$ of the collection optics ($NA = 0.28$), the gray-scale
images exhibit a triangular array of bright, circular spots with approximately 
55\% optical contrast between the bright regions, with maximum intensity 
$I_b$, and the dark regions, with minimum intensity $I_d$. The optical 
contrast is the change in intensity normalized by the average intensity
$2(I_b - I_d)/(I_b +I_d$). The center of a spot corresponds to 
the tip below the center of a channel-glass core. Similar results are seen for 
both $\lambda$. For intermediate $NA$ ($NA = 0.55$), the images for the two $\lambda$ 
are different. For the longer wavelength ($\lambda = 670$ nm), the bright spots
flatten, the light to dark contrast is reduced to 38\%, and the overall
signal level increases. For the shorter 
wavelength ($\lambda = 488$ nm), the bright spots becomes rings, brightest
when the tip is near but not at the channel glass edge, weaker 
when near the center of the channel 
glass core, and weakest outside the channel glass. The light to dark contrast
is reduced to 25\%. At the highest $NA$ ($NA = 0.70$) 
this ring structure persists
for the shorter wavelength with the contrast reduced to 15\%. For the longer 
wavelength this ring structure begins to appear and the 
the contrast ratio is reduced to 28\%. These results are 
independent of the polarization of the tip field and weakly dependent on the
aperture size and the tip/sample separation in the near field. 
Typical line scans are shown
in Fig.~\ref{f2} for $NA = 0.70$ with the tip about 10 nm from the sample. 
The scans were done along $x$, a direction
which goes between two adjacent glass cores, as shown in Fig.~\ref{f1}. Diminished
contrast and a weak contrast reversal are observed when the tip and sample are
widely separated by approximately 1 $\mu$m. 

\section{Theory}

To develop a complete theory for imaging in transmission, we must have
models for the incident tip field and for the electromagnetic modes of the
sample. We must correctly account for the coupling of the incident field
to the sample modes and for the coupling of the excited sample modes to 
reflected and transmitted fields. Finally, we must model the collection of the
transmitted light. These steps are modeled in the following four sections.

\subsection{Tip field}

Rigorous, three-dimensional calculations of the fields emitted by a tip
with a nanometer aperture are difficult and numerically intensive 
\cite {gr6,vl}. The Bethe-Bouwkamp (BB) model \cite {be,bouw} is a simple 
approach that is commonly and successfully used \cite {vl,bc1} 
instead of the numerically intensive approaches to model the tip field. 
In this paper, we use the BB model. As our results will show, the BB
model provides an adequate representation for the tip-fields in the 
experiments we describe so we did not test any of the models (see for 
example \cite {ar1,ar2,grober}) that improve upon the BB model. 

In the Bethe-Bouwkamp model, the
near field of a tapered, metal-coated tip is modeled by the near field
of the light transmitted by a circular aperture with radius $a$, in 
a perfectly conducting, thin,
metal screen when a linearly polarized, plane-wave field is incident 
normal to the screen (see Fig.~\ref{f1}). Bethe and Bouwkamp 
calculated the fields, accurate to order $k_0a$, $k_0 = 2\pi/\lambda$,
for this problem 50 years ago \cite {be,bouw}. Most NSOM experiments 
are done in the range
$0.1 < k_0a < 1.5$, where the BB fields give reasonable results. 
As pointed out by Grober {\it et al.}, the
higher-order corrections to the BB result change only the ratio of
electric to magnetic field amplitudes without strongly affecting the
spatial variation of the field \cite {grober}. 
 
Bethe originally solved the problem in terms of a fictitious magnetic 
surface-charge density, $\rho_m$, and a magnetic surface-current density,
${\bf J}_m$, that exist inside the hole in the screen. Bouwkamp corrected
the ${\bf J}_m$ determined by Bethe. Magnetic vector
and scalar potentials, ${\bf A}_m$ and $\phi_m$ were determined from
${\bf J}_m$ and $\rho_m$. The electric and magnetic fields transmitted by 
the aperture, ${\bf E}_{ap}$ and ${\bf B}_{ap}$ were obtained by taking 
appropriate derivatives of the vector and scalar potentials. Summarizing,
\begin{equation}
{\bf E}_{ap} = \nabla \times {\bf A}_m,
\label{eq_e}
\end{equation}
\begin{equation}
{\bf B}_{ap} = -ik_0{\bf A}_m - \nabla \phi_m,
\label{eq_b}
\end{equation}
where \begin{equation}
{\bf A}_m = - \int {\bf J}_m \frac{e^{ik_0r}}{r}dA, 
\label{eq_vecpot}
\end{equation} \begin{equation} \phi_m = \int \rho_m \frac{e^{ik_0r}}{r}dA.
\label{eq_scalpot}
\end{equation}
The integrals are done over the area $A$ of the aperture, and $r$ is 
the distance from the field point to the source point in the aperture.  
For a plane-wave electric field, linearly polarized along $x$ with amplitude
$E_0$ and incident normal to the screen, the magnetic charge and current 
densities at a source point $(x,y)$ inside the aperture are,
\begin{equation}
\rho_m = - \frac{2yE_0}{\pi^2 \sqrt{a^2 - x^2 - y^2}}
\label{eq_mch} \end{equation}
and \begin{equation}
J_{mx} = - \frac{2ik_0xyE_0}{3\pi^2 \sqrt{a^2 - x^2 - y^2}},
\label{eq_mcurx}
\end{equation}
\begin{equation}
J_{my} = \frac{2ik_0E_0(2a^2 - x^2 - 2y^2)}{3\pi^2 \sqrt{a^2 - x^2 - y^2}},
\label{eq_mcury}
\end{equation}
and
\begin{equation}
J_{mz} = 0.
\label{eq_mcurz}
\end{equation}
The fields transmitted when the incident field is linearly polarized along
$y$ can be obtained from Eqs. \ref{eq_e}-\ref{eq_mcurz} by applying the
appropriate rotations. Transmitted fields for arbitrary incident polarization
can be obtained from the appropriate linear combination of these two sets
of fields.

All of the experimental images were made with the tip axis, taken here
to be along the $z$ direction as shown in Fig.~\ref{f1}, parallel to the glass rods
in the NCG and normal to the NCG surface. 
To model images, we must do calculations for different positions of the
aperture in an $x$-$y$ plane at a fixed distance from the NCG and for different
distances between the tip and the NCG surface. These calculations are most 
easily done in $x$-$y$ Fourier space because the NCG is periodic in $x$ and 
$y$. To do these calculations we need the 2D Fourier
transform of the tip fields. We first obtain the 2D Fourier
transform of the tip fields in an $x$-$y$ plane a distance $z_0$ from the 
plane of the
aperture, ${\bf E}_{ap}(k_x,k_y,z_0)$ and ${\bf B}_{ap}(k_x,k_y,z_0)$, 
by numerical fast Fourier transform. From the free-space propagation, 
we can obtain the Fourier transform of the tip field at any other $x$-$y$ 
plane in free space at a distance z from the aperture. For the electric field:
\begin{equation}
{\bf E}_{ap}(k_x,k_y,z) = \exp (ik_z(z-z_0)){\bf E}_{ap}(k_x,k_y,z_0)
\label{eq_fte}
\end{equation}
where the wavevector $k_z$ for propagation along z is obtained from the
free-space dispersion of a mode with transverse wavevector $(k_x,k_y)$,
$k_z = \sqrt{k_0^2 - k_x^2 - k_y^2}$. The phase of $k_z$ is 
chosen so that the propagating modes ($k_z$ real) are moving forward away from the
aperture and the evanescent modes ($k_z$ imaginary) are damped going away from the
aperture. A similar equation applies 
for the magnetic field. Translations of the tip field in an $x$-$y$ plane
are achieved by translating the phase of the Fourier transform of the field. 
Shifting the tip from $(0,0,z_{ap})$ to $(x_{ap},y_{ap},z_{ap})$ changes the
Fourier components from ${\bf E}_{ap}(k_x,k_y,z_{ap})$ to
$\exp (-i(k_x x_{ap} + k_y y_{ap})){\bf E}_{ap}(k_x,k_y,z_{ap})$.

\subsection{Optical modes of the Nanochannel Glass Array}

Typically, when the modes of a 2D photonic crystal are determined, the 
band structure for the mode frequency, $\omega$ versus $k_x$ and 
$k_y$, is found for $k_z = 0$ \cite{pbs}. To model transmission 
through an NCG, we need a different set of the
optical modes; we need all propagating and evanescent modes, all $k_z$, 
for a given $k_x$, $k_y$, and $\omega$. For that reason, we solve
an eigenvalue problem for $k_z$ rather than an eigenvalue problem for
$\omega$. We begin from the Maxwell equations for electric and magnetic
fields, ${\bf E}$ and ${\bf B}$, with frequency $\omega$,
\begin{eqnarray}
\nabla \cdot {\bf B} & = & 0 \nonumber \\
\nabla \cdot n^2 {\bf E} & = & 0 \nonumber \\
\nabla \times {\bf B} & = & - n^2 i (\frac{\omega}{c}){\bf E} \nonumber \\
\nabla \times {\bf E} & = & i (\frac{\omega}{c}){\bf B},
\end{eqnarray}
where $n$ is the local index of refraction in the NCG. Defining
${\bf F} = n {\bf E}$, then
\begin{eqnarray}
\nabla \times {\bf E} = \nabla \times (\frac{1}{n}{\bf F}) = 
-\frac{\nabla n}{n^2} \times {\bf F} + 
\frac{1}{n} \nabla \times {\bf F}, \nonumber \\
\nabla \cdot n^2 {\bf E} = \nabla \cdot n {\bf F} = 
n \nabla \cdot {\bf F} + {\bf F} \cdot \nabla n.
\end{eqnarray}
From this we have the following Maxwell equations,
\begin{eqnarray}
\nabla \cdot {\bf B} & = & 0 \nonumber \\
\nabla \cdot {\bf F} + {\bf F} \cdot \frac{\nabla n}{n} & = & 0 \nonumber \\ 
\nabla \times {\bf B} & = & - n i (\frac{\omega}{c}){\bf F} \nonumber \\
\nabla \times {\bf F} - \frac{\nabla n}{n} \times {\bf F} & = &
n i (\frac{\omega}{c}) {\bf B}.
\end{eqnarray}
Using the definition ${\bf g} = \nabla n/n$, the Maxwell
equations become,
\begin{eqnarray}
\nabla \cdot {\bf B} & = & 0 \nonumber \\
\nabla \cdot {\bf F} + {\bf F} \cdot {\bf g} & = & 0 \nonumber \\
\nabla \times {\bf B} & = & - n i (\frac{\omega}{c}){\bf F} \nonumber \\
\nabla \times {\bf F} - {\bf g} \times {\bf F} & = &
n i (\frac{\omega}{c}) {\bf B}.
\end{eqnarray}               

To find the modes of the NCG, we treat the 
NCG as translationally invariant along
$z$. The NCG is periodic in $x$ and $y$, so $n$ and ${\bf g}$ have the form
\begin{eqnarray}
n(x,y,z) & = & \sum_{{\bf G}} n({\bf G}) \exp (i(G_xx +G_yy)) \nonumber \\
{\bf g}(x,y,z) & = & \sum_{{\bf G}} {\bf g}({\bf G}) \exp (i(G_xx +G_yy)).
\label{def_n}
\end{eqnarray}
The field modes must have the form,
\begin{eqnarray}
{\bf B}^{m{\bf q}}(x,y,z) & = & \sum_{{\bf G}} {\bf B}^{m{\bf q}}({\bf G}) 
\exp (i((q_x + G_x)x +(q_y + G_y)y + k_z^{m{\bf q}}z)) \nonumber \\
{\bf F}^{m{\bf q}}(x,y,z) & = & \sum_{{\bf G}} {\bf F}^{m{\bf q}}({\bf G})
\exp (i((q_x + G_x)x +(q_y + G_y)y + k_z^{m{\bf q}}z)),
\label{def_e}
\end{eqnarray}
where the ${\bf G}$ are the 2D reciprocal lattice vectors for the NCG
triangular lattice, ${\bf q}$ is in the first Brillouin zone and
$k_z^{m{\bf q}}$ is the eigenvalue for the $m$-th mode for a given ${\bf q}$.

Define $k_x = q_x + G_x$, $k_y = q_y + G_y$ and the convolution
$(n \ast F)({\bf G}) = \sum_{{\bf G}^\prime} 
n({\bf G}-{\bf G}^\prime) {\bf F}({\bf G}^\prime)$
where the sum is over the 2D reciprocal lattice vectors ${\bf G}^\prime$ and similar 
definitions apply for other pairs of $n$ or ${\bf g}$ convolved 
with ${\bf B}$ or ${\bf F}$. For a given ${\bf q}$, the Maxwell
equations couple field Fourier components with different ${\bf G}$.
The equations are (suppressing the Fourier and eigenmode indices),
\begin{eqnarray}
i k_x {\bf B}_y - i k_y {\bf B}_x + i (\frac{\omega}{c}) n \ast {\bf F}_z 
& = & 0 \\
i k_x {\bf F}_y - i k_y {\bf F}_x - g_x \ast {\bf F}_y + g_y \ast {\bf F}_x
 - i  (\frac{\omega}{c}) n \ast {\bf B}_z & = & 0 \\
i k_x {\bf B}_x + i k_y {\bf B}_y + i k_z {\bf B}_z & = & 0 \\
i k_x {\bf F}_x + i k_y {\bf F}_y + i k_z {\bf F}_z + 
g_x \ast {\bf F}_x + g_y \ast {\bf F}_y & = & 0 \\
i k_y {\bf B}_z - i k_z {\bf B}_y & = & -i (\frac{\omega}{c}) n \ast {\bf F}_x \\
i k_z {\bf B}_x - i k_x {\bf B}_z & = & -i (\frac{\omega}{c}) n \ast {\bf F}_y \\
i k_y {\bf F}_z - i k_z {\bf F}_y - g_y \ast {\bf F}_z & = & 
i (\frac{\omega}{c}) n \ast {\bf B}_x \\
i k_z {\bf F}_x - i k_x {\bf F}_z + g_x \ast {\bf F}_z & = &
i (\frac{\omega}{c}) n \ast {\bf B}_y.
\end{eqnarray}
The first two equations, Eqs. (16) and (17), come from equations that have no derivatives of
the $z$ field components and have no dependence on $k_z$. These equations provide
homogeneous constraints on the solutions to the eigenvalue problem defined by
Eqs. (18)-(23). Written in explicit eigenvalue form, Eqs. (18)-(23) become,
\begin{eqnarray}
k_x {\bf B}_z - (\frac{\omega}{c}) n \ast {\bf F}_y & = & k_z {\bf B}_x \\
k_y {\bf B}_z + (\frac{\omega}{c}) n \ast {\bf F}_x & = & k_z {\bf B}_y \\
- k_x {\bf B}_x - k_y {\bf B}_y & = & k_z {\bf B}_z \\
(\frac{\omega}{c}) n \ast {\bf B}_y + k_x {\bf F}_z + i g_x \ast {\bf F}_z 
& = & k_z {\bf F}_x \\
- (\frac{\omega}{c}) n \ast {\bf B}_x + k_y {\bf F}_z + i g_y \ast {\bf F}_z 
& = & k_z {\bf F}_y \\
- k_x {\bf F}_x + i g_x \ast {\bf F}_x - k_y {\bf F}_y + i g_y \ast {\bf F}_y 
& = & k_z {\bf F}_z.
\end{eqnarray}
Eqs. (24)-(29) and the two constraints, Eqs. (16) and (17), guarantee that for a
given ${\bf q}$ we will have four modes per ${\bf G}$. These four
modes correspond to two polarizations for each $k_z$ and  
$- k_z$. For $k_z \rightarrow - k_z$, we have $({\bf F}_x,{\bf F}_y,{\bf F}_z) \rightarrow
(-{\bf F}_x,-{\bf F}_y,{\bf F}_z)$ and $({\bf B}_x,{\bf B}_y,{\bf B}_z) \rightarrow
({\bf B}_x,{\bf B}_y,-{\bf B}_z)$, as can be checked by inspection of Eqs. (16), (17),
and (24)-(29). 

\subsection{Field transmission}

We determine the transmission of the incident tip field 
through the NCG by solving the 
standard boundary matching problem for transmission through a film.
We assume that the total field in the free-space region
containing the tip is the incident tip field plus
the field reflected by the NCG. Explicitly, the reflected field is
\begin{equation} 
{\bf E}_r(x,y,z) = \sum_{{\bf k},p} {\bf E}_{r,p}({\bf k})
\exp(i(k_xx + k_yy + k_z^rz)),
\end{equation}
where the sum is over lateral wavevectors ${\bf k} = (k_x,k_y)$ 
and the two polarizations $p$. 
The $z$ wavevector $k_z^r$ is obtained from the free-space dispersion 
$k_z^r = \sqrt{k_0^2 - k_x^2 - k_y^2}$. The phase of $k_z^r$ is 
chosen so that propagating modes are forward propagating and evanescent modes are damped 
as the mode moves from the NCG sample back toward the tip. The two polarization
vectors ${\bf E}_{r,p}({\bf k})$ are chosen to ensure that $\nabla \cdot {\bf E}
= 0$ for each free-space mode. Reflections off
the tip are not included. This is the only approximation, other than the use
of the Bethe-Bouwkamp model for the source field, that we make for the local fields
near the tip. As our results will show, including multiple reflections is 
not important for the experiments we model.
${\bf B}_r$ is defined by a similar equation. The field in the free-space region
containing the collection optics is the transmitted field,
\begin{equation}
{\bf E}_t(x,y,z) = \sum_{{\bf k},p} {\bf E}_{t,p}({\bf k})
\exp(i(k_xx + k_yy + k_z^tz)),
\end{equation}
where $k_z^t$ is determined from the free-space dispersion. The phase of $k_z^t$ is
chosen so that propagating modes are forward propagating and evanescent modes are damped
as the mode moves from the NCG sample toward the collector.
The field in the NCG is
\begin{equation}
{\bf E}_{ncg}(x,y,z) = \sum_{m,{\bf q}} \alpha_{m{\bf q}}
{\bf E}^{m{\bf q}}(x,y,z),
\end{equation}
where the $\alpha_{m{\bf q}}$ are the amplitudes for the modes 
excited in the NCG. 
At each NCG surface, we have the boundary conditions that the two tangential
electric field components, $E_x$ and $E_y$, and the two tangential magnetic
field components, $B_x$ and $B_y$, be continuous. These four conditions 
define the set of matrix equations that are solved to determine
the amplitudes for the Fourier components of the reflected and
transmitted fields and for the excited NCG modes.

When calculating transmission through a film, one must worry about the
effects of multiple reflections between the sample surfaces that can
produce resonant transmission and Fabry Perot oscillations.
We calculate transmission for thick NCG samples, as used in the
experiments. These samples are much thicker than $\lambda$. We also do calculations
for films much thinner than $\lambda$. We study these two limits to investigate the
effect of sample thickness on image formation. We have no problem doing the
calculations when the film is thinner than $\lambda$, because 
transmission resonances do not occur. For the thick samples, we have to be
more careful. Resonances can be sharp when the sample has a uniform thickness 
that is much greater than $\lambda$. In real samples these resonances are
broadened because the sample thickness is not uniform. However, our 
calculations are done for samples with uniform thickness. Spurious effects 
from sharp resonances in the transmission can occur because we 
include only a finite set of reciprocal 
lattice vectors in our representations of the index and the 
fields (Eqs. \ref{def_n} 
and \ref{def_e}). Typically, we sum over all reciprocal lattice
vectors with lengths less than 3-4 times the length of the fundamental
reciprocal lattice vectors. Numerical diagonalization to find the eigenmodes becomes
very time consuming if we include more reciprocal lattice vectors.
Small changes in the cutoffs make 
only small changes in the modes. This can result in large changes of the
specific modes that are resonant at a particular sample thickness. 
We have checked that the effects of
these resonances are spurious by varying cutoffs for the wavevectors we
include and by varying the sample thickness. We can eliminate
these spurious effects by increasing the cutoffs. However, for
typical cases, it is computationally too expensive to increase the
cutoff enough to eliminate all spurious effects. To eliminate these
spurious effects simply, we add a damping factor to all of the
propagating NCG modes. For all NCG modes with real $k_z$, we include
an imaginary term (typically 0.01 $(\mu m)^{-1}$) to the
$z$ wavevector to damp waves as they propagate across the sample.
This damping reduces the effects of the multiple reflections between the sample
surfaces, broadens the resonances, and thereby eliminates the spurious effects 
in the contribution from resonances. In the experiments, variations in
sample thickness break the coherence of multiple reflections and
eliminate resonance effects. Our use of a damping factor is a simple way
to mimic this phase breaking and eliminate resonance effects.
We use the same damping for all propagating waves. We also choose
this damping to be much weaker than the damping of any of the evanescent
modes. As a consequence, the damping we introduce reduces the overall intensity
of the transmitted light without affecting the relative transmission amplitudes 
of different Fourier components of the transmitted light. Thus, the damping
that we include does not affect the transmission line scans we calculate, other than
to change the absolute scale of the transmission intensities. We get the
same transmission line scans, except for the overall scale factor, when we eliminate
spurious effects of the resonances by increasing the wavevector cutoff. 

\subsection{Image formation}

To model transmission-NSOM images, we need to calculate the amount of
the transmitted light that is collected by optics with a numerical aperture
$NA$. We use the following simple model (see Fig.~\ref{f1}). We assume that the optics
collects all of the transmitted flux that passes through an $x$-$y$ plane in the
far-field away from the NCG. All flux leaving the NCG within the numerical
aperture of the optics, the angle $\theta$ in Fig.~\ref{f1}, is collected.

To calculate the total intensity of the collected light, $I$, we first need the 
time-averaged Poynting vector for a harmonic free-space field with frequency $\omega$,
\begin{equation}
\langle {\bf S}(x,y,z) \rangle = \frac{c}{8 \pi} Re \{{\bf E}(x,y,z) \times {\bf B}^{\ast}(x,y,z)\}.
\end{equation}
The flux passing through an $x$-$y$ plane in the far-field is 
$\langle S_z(x,y,z \rightarrow \infty ) \rangle$.

Using a 2D Fourier expansion to write the fields,
\begin{equation}
{\bf E}(x,y,z) = \sum_{k_x,k_y} {\bf E}(k_x,k_y) \exp (i(k_xx +k_yy)) \exp (ik_z(z-z_s)),
\end{equation}
where $z_s$ is the position of the back sample-surface, the sum is over all 2D $k$-space,
$k_z$ is obtained from the free-space dispersion, and the phase of $k_z$ is 
chosen to give forward-propagating or damped waves 
going into free space away from $z_s$. A similar equation holds
for ${\bf B}$. The integrated flux passing through the $x$-$y$ plane at $z$ is
the total intensity,
\begin{eqnarray}
I_{tot} & \equiv & \frac{1}{(2 \pi)^2} \int dx\,dy\, 
\langle S_z(x,y,z) \rangle \nonumber \\
& = &
\frac{c}{8 \pi} Re\{\sum_{k_x,k_y} \hat{\bf z} \cdot ({\bf E}(k_x,k_y) \times {\bf B}
^{\ast}(k_x,k_y))\exp(i(k_z - k_z^{\ast})(z-z_s))\}.
\end{eqnarray}
For propagating modes $k_z = k_z^{\ast}$; for evanescent modes $k_z = - k_z^{\ast}$.
In the far-field limit, the exponential factor vanishes for evanescent modes and is 1 for
propagating modes, thus 
\begin{equation}
I_{tot} = \frac{c}{8 \pi}\sum_{\sqrt{k_x^2 + k_y^2} \leq \frac{2\pi}{\lambda}} Re 
\{ E_x(k_x,k_y)B_y^{\ast}(k_x,k_y) - E_y(k_x,k_y)B_x^{\ast}(k_x,k_y)\}.
\end{equation}
The optics collects light within the cone defined by the numerical aperture,
reducing the cutoff for light collection from the free-space cutoff
$2\pi/\lambda$ to $2\pi NA /\lambda$ where $NA = \sin \theta$ in free space.
Thus, the total intensity of transmitted light $I$ collected by the optics is 
\begin{equation}
I = \frac{c}{8 \pi}\sum_{\sqrt{k_x^2 + k_y^2} \leq \frac{2\pi NA}{\lambda}} Re
\{ E_x(k_x,k_y)B_y^{\ast}(k_x,k_y) - E_y(k_x,k_y)B_x^{\ast}(k_x,k_y)\}.
\label{I_final}
\end{equation}

Simulated images are obtained by calculating $I$ as a function of the relative
positions of the tip and sample. In this paper we present one-dimensional line
scans of these images. 

\subsection{Nanochannel glass model}

As shown in Fig.~\ref{f1}, the NCG sample is a 2D triangular lattice of glass rods in
a glass matrix. We assume that rods are perfect cylinders. In the experiments,
the lattice spacing is 1.07 $\mu$m and the core radius $r_c = 0.37$ $\mu$m. 
We use these values to model the measured images. We do additional calculations
with other lattice parameters to  
further test the potential of NSOM for probing these structures. The measured 
samples are approximately 250 $\mu$m thick. We use this thickness to model
the measured images. Our results for thick samples 
are insensitive to sample thickness because
we eliminate Fabry-Perot oscillations by use of the damping factor.
We also consider thin samples where sample evanescent
modes, excited by the tip field, can be transmitted through the sample.
We assume that the index of refraction is real and uniform in each glass, 
has the bulk value in each glass, and has a step discontinuity at
the core/matrix glass interface. This is the only assumption that we make
about the dielectric structure of the NCG. The matrix 
glass has a bulk index of refraction of 1.678 for $\lambda = 488$ nm and
1.657 for $\lambda = 670$ nm \cite{hsu}.
The core glass has a higher index so it acts as a guiding region. For
$\lambda = 488$ nm, the index difference is $\Delta n \simeq 0.011$ \cite{hsu}. For
$\lambda = 670$ nm, the index difference is $\Delta n \simeq 0.019$ \cite{hsu}. We use
these values to model the measured images. Other calculations are done with 
higher $\Delta n$ for additional tests. The scanned NCG-surface 
was polished. The core glass etches preferentially during polishing, leaving 
$3.5 \pm 0.5$ nm depressions centered on the channel glasses. The experimental
scans are done at constant separation, $z_{ap}$, 
between the tip aperture and the NCG. For 
the calculations, we model a flat surface, ignoring the effect of these 
small depressions on the fields at the sample surface. Our results show that
the transmission NSOM images are determined by the coupling of the 
tip fields to the {\it bulk} photonic modes of the NCG and should not depend on the surface
structure. Also, our results do not change much for nm-scale changes in $z_{ap}$ or 
sample thickness so we can ignore topographic effects on the calculated images.

\subsection{Significance of $NA$ and $k$-space cutoffs}

Our NCG experiments \cite{hsu}, as well as other experiments \cite{hsu2},
show that collection optics $NA$ has a strong effect on transmission
NSOM image contrast. The collection optics $NA$ determines which transmitted 
modes are collected.
Figure~\ref{f3} shows the reciprocal-lattice points for the sample
studied experimentally. Also shown are the boundary for the
first Brillouin zone and the cutoffs for the transverse wavevector
that determine which transmitted modes are collected
by the optics for different $\lambda$ and $NA$ (see Eq. \ref{I_final}). 
The transverse wavevectors for transmitted
modes that are collected by the optics
lie inside the cutoff. As $NA$
increases, the cutoff for mode components that are collected by
the far-field detector increases.  Also, for fixed $NA$ the cutoff increases
with decreasing $\lambda$. For $NA = 0.28$, the cutoffs lie well inside
(at) the boundary of the first Brillouin zone for $\lambda = 
670$ nm (488 nm). For $NA = 0.7$, the cutoff extends to the 
edge of the second Brillouin zone for $\lambda =
670$ nm and into the fourth Brillouin zone for $\lambda =
488$ nm. From our results we find that the scans for 
different $\lambda$ are similar if the scans have similar cutoffs and thus
have contributions from similar sets of transverse wavevectors. Thus the scans
for $\lambda = 670$ nm  are similar to those scans for $\lambda = 488$ nm
done at smaller $NA$.

\section{Results}

\subsection{Line scans}

First, we show that the theory is able to describe the images of the 
NCG made with 
transmission NSOM. Calculated line scans along $x$ of the transmitted
intensity are shown in Figs.~\ref{f4} and \ref{f5} for $\lambda = 488$ nm and 670 nm 
respectively.
The intensities are given in arbitrary units. However, the same units are used
for these two figures and for all other figures that deal with the same
NCG, except where it is noted that the results have been rescaled. Thus,
intensities plotted in different figures for the same NCG can usually be 
compared directly.
The dependence on $NA$ is shown. The calculated line scans reproduce the
key features in the measured line scans. For small $NA$, the line scans
are peaked at the channel glass centers, with a peak/valley intensity ratio
of about 1.5, even though the index contrast is only 1-2\%. As $NA$ increases,
more of the transmitted light is collected and the average transmitted 
intensity increases.
The peaks broaden and flatten and the peak/valley ratio decreases. At $NA =
0.5$ $(0.7)$ for $\lambda = 488$ $(670)$ nm, the peaks about core centers develop 
structure with depressions in the middle and side peaks $\approx 0.2$
$\mu$m from the centers. These side peaks produce the ring structure
observed in the experimental images. The experimental images at 0.7 $NA$ 
show the rings prominently for $\lambda = 488$ nm but only hint that rings are
starting to appear for $\lambda = 670$ nm. Our calculations also show that the rings
form at lower $NA$ for shorter $\lambda$. This occurs because 
more of $k$-space is collected, for a given $NA$,  for shorter $\lambda$. Scans along
other directions (see Fig.~\ref{f6} for scans along $y$) give similar results. 
In particular, the structure inside the glass core is identical for the 
scans along different directions, confirming the cylindrical symmetry of this
structure. This gives evidence that the structure in the peak 
is due to transmitted light that propagates through the sample in photonic modes that are
concentrated in the cylindrically symmetric cores.
 
Figure~\ref{f2} shows a comparison of the experimental and calculated line scans
along $x$ at $NA = 0.7$ for both $\lambda$. The curves have been scaled to have the
same average value and then 
shifted for clarity. There is strong qualitative agreement between
the data and the calculations, as we have already discussed. The
ring structure occurs at similar positions, well inside the edge
of the channel glass, in both the experimental
and calculated scans. This ring structure
is not directly related to the channel glass edge but is connected to
the spatial distribution of NCG photonic modes. The most noticeable 
differences are in the intensity ratios. The depressions in the peaks for 488 nm are
more pronounced in the experimental line scans than in the calculated scans. The
experimental data typically shows more contrast than the calculations. For example,
the small dip in intensity at the three-fold site for a $y$ scan is harder to see in the
calculated scans (see Fig.~\ref{f6}) than in the experimental data \cite {hsu}.
The experimental intensity ratios at $0.28 NA$ for peak/three-fold site and
bridge/three-fold site  
are 1.58 and 1.18 for $\lambda = 488$ nm and 1.60 and 1.10 for
$\lambda = 670$ nm \cite{hsu}. The calculated ratios at $0.3 NA$ are 1.69 and 1.05 
for $\lambda = 488$ nm and 1.51 and
1.02 for $\lambda = 670$ nm. At $0.7 NA$ the experimental peak/three-fold site 
intensity ratios are 1.15 for $\lambda = 488$ nm and 1.28 for 
$\lambda = 670$ nm \cite{hsu}. The calculated ratios at $0.7 NA$ are 1.08
for $\lambda = 488$ nm and 1.14 for $\lambda = 670$ nm. In all cases, the experimental and
calculated intensity ratios are similar. In all
cases except one, the experimental ratio is larger than the calculated ratio. This suggests
that the calculations may underestimate slightly the coupling to the NCG
modes that are concentrated in the glass cores and
produce the structure in the line scans. Some of the small disagreement
between experimental and calculated intensity ratios can probably be removed by
using better models for the NCG and its dielectric profile, 
the tip field, or the multiple scattering
between the tip and the NCG. We have not yet tested these possibilities.

Consider the following {\it simple} analysis of the transmitted intensity to understand
the structure in the line scans. 
The index difference between the core and matrix glasses in the NCG is small, so much of the
collected light will be transmitted through the NCG in extended, nearly uniformly 
distributed modes. The channel glass does have the higher index, so some of
the collected light will be transmitted in modes more concentrated in the glass cores.
The line scans can be better understood by separating the
collected intensity $I$ into these two contributions, 
$I = I_m + \Delta I$. $I_m$ is the minimum intensity. This is the uniform intensity that is
transmitted and collected, regardless of the tip position. 
$\Delta I$ is the variation in the line scan above the
minimum intensity. $\Delta I$ includes the structure in the
line scan. Figs.~\ref{f4}-\ref{f6} show that $I_m$ is 
approximately the collected intensity at the bridge site, which 
varies approximately as $NA^2$. The uniform response of the NCG is proportional
to the total phase space area of the collected light. The amplitude for 
variations of $\Delta I$ for a line scan along $x$ is the difference in intensity,
$I_c -I_b$, between center and bridge sites. As a function of increasing
$NA$, $I_c -I_b$ first rapidly increases, peaks, and then slowly decreases.
The amplitude for variations in $\Delta I$ is a maximum for line scans collected
with $NA \approx 0.3-0.35$ for $\lambda = 488$ nm and 
$NA \approx 0.4-0.45$ for
$\lambda = 670$~nm. For both $\lambda$, the maximum amplitude for variations in
a line scan occurs for collection with $\lambda/NA \approx 4 r_c$, where $r_c$ 
is the radius of the channel glass. The minimum wavelength, $\lambda_{min}$, 
for the in-plane variation of the 
light collected by the optics for a particular $NA$ is $\lambda/NA$. The maximum
amplitude for variations in a line scans occur when the scans are collected with an 
$NA$ that 
just accepts light concentrated in modes that can fill the channel glass core with 
half a wavelength (i.e. the lowest order ``channel" mode with transverse wavelength
$\lambda_{trans} \approx 4 r_c \approx \lambda_{min}$, as one would expect if the
channels were waveguides with strong confinement of the fields).

\subsection{Importance of the tip field}

Calculations have been done to determine the effect of structure in the tip field
on the line scans. In the Bethe-Bouwkamp model for the aperture field, the polarization
of ${\bf E}_{ap}$ is determined by the polarization of the plane-wave field incident
on the hole in the screen. We tried different polarizations of the incident field 
in the BB model and found no significant differences in the calculated line scans.
This is consistent with the experimental observation that the images were insensitive
to polarization. This insensitivity is another indication that the local variations 
in the images arise from the NCG photonic modes that are concentrated in the
channel-glass cores and have cylindrical symmetry in the cores.

The effect of the separation, $z_{ap}$, of the aperture from the NCG and of the aperture
radius, $a$, on the line scans has been tested as well. 
Figure~\ref{f7} shows the dependence on $z_{ap}$
for $\lambda = 488$ nm and $0.7 NA$. Similar results were seen for other 
$NA$s and for $\lambda = 670$ nm.
As the tip/sample separation increases, there is a steady decrease in the 
total collected intensity but little change in the structure in $I$ until the
intensity begins to increase and the contrast reverses at large $z_{ap}$ ($z_{ap}
\approx 500$ nm). This suggests that the NCG modes that transmit
light through the thick film do not couple efficiently to the strongly-localized evanescent
modes emitted by the tip and that the structure in the line scans is related to the mode
density in the NCG rather than the field distributions near the tip. Line scans 
calculated by including only the propagating 
tip modes are almost identical to line scans calculated with both the propagating and
the evanescent tip modes. This confirms the experimental observation that the 
evanescent tip modes did not couple effectively to the modes in the NCG that was imaged. 
$I$ increases at large $z_{ap}$ because 
the tip field spreads laterally enough to couple efficiently 
to multiple channel cores at the same time. When $z_{ap}$ is small, the tip field couples 
to one core at a time. The contrast reverses at large $z_{ap}$ possibly because the
coupling to multiple channels simultaneously is more efficient when the tip is between
channels rather than centered on a channel. Contrast reversal at large $z_{ap}$ was
also observed experimentally.

The dependence of the line scans on tip radius $a$ is shown in Fig.~\ref{f8} at 0.7 $NA$
and $\lambda = 488$ nm for the tip close to the NCG, {\it i.e.} $z_{ap} = 10 $ nm.
Similar results were seen for other $NA$ and $\lambda$. The intensity of the
aperture field increases rapidly with increasing $a$. We eliminate this dependence by 
scaling the line scans to have the same average intensity. In the figure, the scans 
have been
shifted for clarity. For small tips with $a \stackrel{<}{\sim} 100$ nm, the structure
is independent of tip size. Again, this suggests that the NCG modes that transmit
light through the thick film do not efficiently couple to the strongly localized tip modes.
The structure persists but weakens for 
100 nm $\stackrel{<}{\sim} a \stackrel{<}{\sim} 300$ nm and 
the contrast reverses for $a \geq 300$ nm. The structure weakens and reverses contrast
when the tip becomes wide enough to couple to multiple channels simultaneously.

\subsection{Significance of photonic crystal optical modes}

In this section we show 
that the structure in the line scans is related more closely to the
density of the photonic modes in the NCG that are concentrated in the cores and not 
as closely to the density of the more extended, nearly uniform modes or to 
the structure in the tip field. We present line scans for
different channel-glass sizes and different NCG lattice constants in Figs.~\ref{f9} and
\ref{f10}. The calculations are done for 0.7 $NA$ and $\lambda = 488$ nm 
for a 50 nm tip radius.
As the size of the channel glass increases, the width of the central peak in the
line scan increases. The separation between the side peaks that produce the
rings in the images also increases. For each core size, 
the ring structure occurs well inside the
channel glass and does not give a direct measure of the channel-glass size. 
Additional structure develops as the glass core increases in size and additional modes
can be concentrated at the cores. 
The ring structure is eliminated by reducing the
core size and making it difficult for any mode except the lowest order mode 
to be concentrated on the cores. The line scans show a similar
dropoff near the glass-core edge for different core sizes, indicating that 
the spread of core-like modes into the matrix glass, which
depends weakly on core-glass size, determines the dropoff. The core-glass size
and the lattice constant can be varied separately. Fig.~\ref{f10} shows that
the structure in the line scans is determined by the NCG modes that are concentrated to the
glass cores. Increasing the NCG lattice constant increases the spacing between the main
features in the scan without changing the structure in those features. 

\subsection{High-contrast samples}

The structure in the experimental images and in the line scans discussed so far results
because the NCG supports photonic modes which are concentrated on the channel-glass cores.
The structure is insensitive to the tip field when the tip field only couples to one
channel at a time. Contrast reversal occurs when the tip field couples to multiple
channels simultaneously. Transmission of the evanescent near fields from the tip 
through the NCG is negligible in the samples studied for two reasons. First, the transverse 
wavevector $k_{ev}$ for localized evanescent 
modes emitted by the tip is much larger than the reciprocal-lattice basis vectors of the
NCG samples, {\it i.e.} $k_{ev} \approx \pi/(2a) \gg G^0_x, G^0_y$. The evanescent modes
can couple to the NCG photonic modes in the experimental samples only through very high-order
Fourier components of the index-of-refraction profile. To enhance 
this coupling, one can reduce the NCG lattice constant so that 
the tip field and the index profile vary on the same length scale in the plane,
{\it i.e.} so that $k_{ev}$ is comparable to $G^0_x$ and $G^0_y$.  One can also
increase the index contrast $\Delta n$. In Fig.~\ref{f11} we show line scans calculated
for an NCG with half the lattice constant and half the core-glass radius of the 
experimental samples, with the same index for the matrix glass as in the experimental
samples, but an index contrast $\Delta n = 1$ which is much bigger than in
the experimental samples. The sample thickness was chosen to be 250 $\mu$m as in the
experimental samples. The calculations were done as a function of $NA$ for 
$\lambda = 488$ nm with a 
50 nm radius tip, scanning 10 nm from the NCG. The structure in the line scans 
is similar to that obtained
for the experimental samples with only 1-2 per cent index contrast. 
At low $NA$, a single central
peak occurs. As $NA$ increases, this peak broadens, flattens and then evolves into 
the ring structure. However, the peak/bridge site contrast ratios are much larger 
when $\Delta n = 1$. A comparison of line scans of
the high $\Delta n$ sample done at $z_{ap} = 10$ nm with line scans done at $z_{ap} = 50$ nm
(see Fig.~\ref{f12}) shows a much stronger dependence on $z_{ap}$ than seen for low
$\Delta n$ samples (Fig.~\ref{f7}). There is a much larger decrease in total transmitted
intensity for the same change in tip height when $\Delta n = 1$. 
More importantly, the structure in the line scans of the high $\Delta n$ sample 
changes significantly as the tip height changes from 10 nm
to 50 nm. The ring structure disappears when $z_{ap}$ is increased from 10 nm to 50 nm 
above the high $\Delta n$ sample. 

These calculations were done for a 250 $\mu$m thick sample. Many of the NCG modes that
couple to tip evanescent fields are evanescent modes. These NCG
modes make a negligible contribution to transmission through a thick
sample. Transmission of evanescent fields
from the tip through the sample can be enhanced by reducing the sample thickness. 
Figure~\ref{f13} shows line scans done for the same sample as for Fig.~\ref{f11} 
except that the sample thickness is only 0.1 $\mu$m. Increased transmission
of the evanescent tip fields by the evanescent sample modes greatly 
enhances the structure in the line scans.
The ring structure is much more prominent. Additional structures at interstitial
sites, such as the bridge sites, is now present. A comparison (Fig.~\ref{f14}) of the line
scans for $z_{ap} = 10$ nm and $z_{ap} = 50$ again shows a strong $z_{ap}$ dependence
both at high and low $NA$. Calculations for samples which are approximately a wavelength
thick show additional Fabry Perot structure. The thickness of the sample determines which
sample modes provide enhanced transmission. The tip/sample separation determines 
how strongly these modes are excited. 

\section{Conclusions}

A theory has been presented to model recent transmission NSOM images made of
nanochannel-glass arrays. The theory describes the entire process of image
formation, including the tip field that is the light source, the photonic
modes of the nanochannel-glass array, coupling of the tip field to the
sample modes, and transmission of the fields to the collection optics.
The theory is able to reproduce the key features observed in the experimental
images, including the evolution of the images with increasing numerical
aperture of the collection optics and the dependence on wavelength. Most importantly,
the theory is able to explain a ring-like structure that develops in the
images as the numerical aperture is increased. 

For the samples studied experimentally, transmission NSOM probes most directly 
the photonic mode structure of the nanochannel-glass array. The structure
in the images can be related to the photonic modes of the array that are 
strongly concentrated at the glass cores. 
The numerical aperture of the collection optics can be varied to control 
the size of the phase space of the
photonic modes that is sampled by the transmission experiments.
Angle-resolved mapping of the photonic modes could be achieved by collecting
light transmitted at specific angles. 
The structure in the images is insensitive to the details of the tip
field because the samples do not couple strongly to the localized evanescent
modes emitted by the tip. In these experiments the images become sensitive
to the tip field only if the tip field is so broad that it can simultaneously
excite multiple channel-glass cores in the array.

Contrast and structure in the images can be enhanced greatly by increasing the
transmission of the evanescent tip fields through the sample. This can be accomplished
by choosing samples with higher index contrast or with index profiles with 
spatial variations that more closely match the spatial localization of the
tip field. This enhanced contrast comes at the price of increased sensitivity to 
the tip field. This increased sensitivity can be exploited to better probe the sample,
but it must be understood to properly separate sample properties from probe properties.

 \acknowledgments{We thank P. S. Julienne for helpful discussions. E. B. McDaniel
 acknowledges financial support from ONR and J. W. P. Hsu from Sloan Research. 
Work done at the University of Virginia was supported by the NSF, and 
that at NRL was supported by DARPA and ONR.}

%\newpage

\begin{figure}
\caption{Schematic of the transmission NSOM experiment. The nanochannel-glass
array sample is scanned at constant separation across the NSOM tip. The transmitted
light is collected by an objective with numerical aperture $NA$. Key elements of
the theory are represented schematically. The tip field is described by the 
Bethe-Bouwkamp model {\protect \cite {be,bouw}}. The collected light is 
modeled as the total flux that 
leaves the sample inside the acceptance cone defined by $NA$ and passes 
through a plane, as indicated, in the far field. The array geometry is
indicated.}
\label{f1}
\end{figure}

\begin{figure}
\caption{Transmission as the tip scans along $x$: experimental scans
($\Box$,$+$) and calculated scans ({\protect\rule[.04in]{.20in}{.01in}}) scaled 
to have the same average intensity and shifted for clarity. The core-glass centers 
are marked by vertical dashed lines. At $x = 0$ the tip and core
glass centers are aligned.} 
\label{f2}
\end{figure}

\begin{figure}
\caption{Reciprocal lattice points ($\bullet$) and 
first Brillouin zone boundary ({\protect\rule[.03in]{.20in}{.03in}}) for
the NCG sample. Cutoff wavevectors for $\lambda = 488$ nm 
({\protect\rule[.04in]{.20in}{.01in}}) and 670 nm ($\cdot\cdot\cdot\cdot$) are
shown. Each inner (outer) circle is for $NA = 0.28$ (0.7).}
\label{f3}
\end{figure}

\begin{figure}
\caption{Transmitted intensity line scan along $x$: $\lambda = 488$ nm,
tip/sample separation $z_{ap} = 10$~nm, tip radius $a = 50$ nm. Dependence on
collection optics $NA$ is shown. Positions of channel-glass centers (c) and 
bridge sites (b), the midpoints on the lines joining adjacent glass centers, are 
shown.}
\label{f4}
\end{figure}

\begin{figure}
\caption{Transmitted intensity line scan along $x$: $\lambda = 670$ nm,
$z_{ap} = 10$ nm, $a = 50$ nm. Dependence on
$NA$ is shown. Positions of channel-glass centers (c) and
bridge sites (b) are shown.}
\label{f5}
\end{figure}

\begin{figure}
\caption{Comparison of transmitted intensity line scans along
$x$ ({\protect\rule[.04in]{.20in}{.01in}}) and $y$ ($\cdot\cdot\cdot\cdot$):
$\lambda = 488$ nm, $z_{ap} = 10$ nm, $a = 50$ nm. Positions of channel-glass 
centers (c), bridge sites (b) along $x$, and 3-fold sites
(3f)along $y$ are shown. The 3-fold symmetric sites are at the middle of the 
triangles formed by 3 adjacent nearest-neighbor glass centers.}
\label{f6}
\end{figure}

\begin{figure}
\caption{Transmitted intensity line scan along $x$: $\lambda = 488$ nm,
$a = 50$ nm, $NA = 0.7$. Dependence on tip/sample separation, $z_{ap}$,
is shown.}
\label{f7}
\end{figure}

\begin{figure}
\caption{Transmitted intensity line scan along $x$: $\lambda = 488$ nm,
$z_{ap} = 10$ nm, $NA = 0.7$. Dependence on tip radius, $a$,
is shown. The line scans have been scaled to have the same average intensity
(1000 on the scale used)
and then shifted for clarity from the scan for $a = 100$ nm by multiples of
100.}
\label{f8}
\end{figure}

\begin{figure}
\caption{Transmitted intensity line scan along $x$: $\lambda = 488$ nm,
$z_{ap} = 10$ nm, $NA = 0.7$, $a = 50$~nm. Dependence on channel-glass 
core radius is shown.}
\label{f9}
\end{figure}

\begin{figure}
\caption{Transmitted intensity line scan along $x$: $\lambda = 488$ nm,
$z_{ap} = 10$ nm, $NA = 0.7$, $a = 50$~nm. Dependence on lattice constant 
and channel-glass core radius is shown. The curves are all on the same
scale but have been shifted for clarity by from the scan for
the experimental sample (lattice
constant 1.07 $\mu$m and core radius $r_c = 0.37$ $\mu$m) by multiples of 300. 
The positions of the core centers
(c) and bridge sites (b) for the NCG studied experimentally are indicated.}
\label{f10}
\end{figure}

\begin{figure}
\caption{Transmitted intensity line scan along $x$: $\lambda = 488$ nm,
$z_{ap} = 10$ nm, $a = 50$ nm. Dependence on $NA$ is shown. 
The NCG lattice constant is 0.535 $\mu$m, the channel-glass core radius
is 0.186 $\mu$m, and the index contrast is $\Delta n = 1.0$. The NCG sample 
is 250 $\mu$m thick. The vertical lines indicate channel-glass centers (c),
bridge sites (b), and edges of the channel glass (e).}
\label{f11}
\end{figure}

\begin{figure}
\caption{Comparison of transmitted intensity line scan along $x$: 
$\lambda = 488$ nm, $a = 50$ nm, $z_{ap} = 10$ nm 
({\protect\rule[.04in]{.20in}{.01in}})
and 50 nm ($\cdot\cdot\cdot\cdot$). 
The NCG has the same lattice parameters, thickness, and
index contrast as in Fig.~\ref{f11}. Dependence on $NA$ is shown.}
\label{f12}
\end{figure}

\begin{figure}
\caption{Transmitted intensity line scan along $x$: $\lambda = 488$ nm,
$z_{ap} = 10$ nm, $a = 50$ nm. Dependence on $NA$ is shown.
The NCG lattice constant is 0.535 $\mu$m, the channel-glass core radius
is 0.186 $\mu$m, and the index contrast is $\Delta n = 1.0$. The NCG sample
is 0.10 $\mu$m thick.}
\label{f13}
\end{figure}

\begin{figure}
\caption{Comparison of transmitted intensity line scan along $x$:
$\lambda = 488$ nm, $a = 50$ nm, $z_{ap} = 10$ nm
({\protect\rule[.04in]{.20in}{.01in}})
and 50 nm ($\cdot\cdot\cdot\cdot$).
The NCG has the same lattice parameters, thickness, and
index contrast as in Fig.~\ref{f13}. Dependence on $NA$ is shown.}
\label{f14}
\end{figure}

%\end{multicols}


\begin{thebibliography}{99}
\bibitem{na} In free space the numerical aperture $NA = \sin(\theta)$ where
$\theta$ is the half angle of the acceptance cone.
\end{thebibliography}
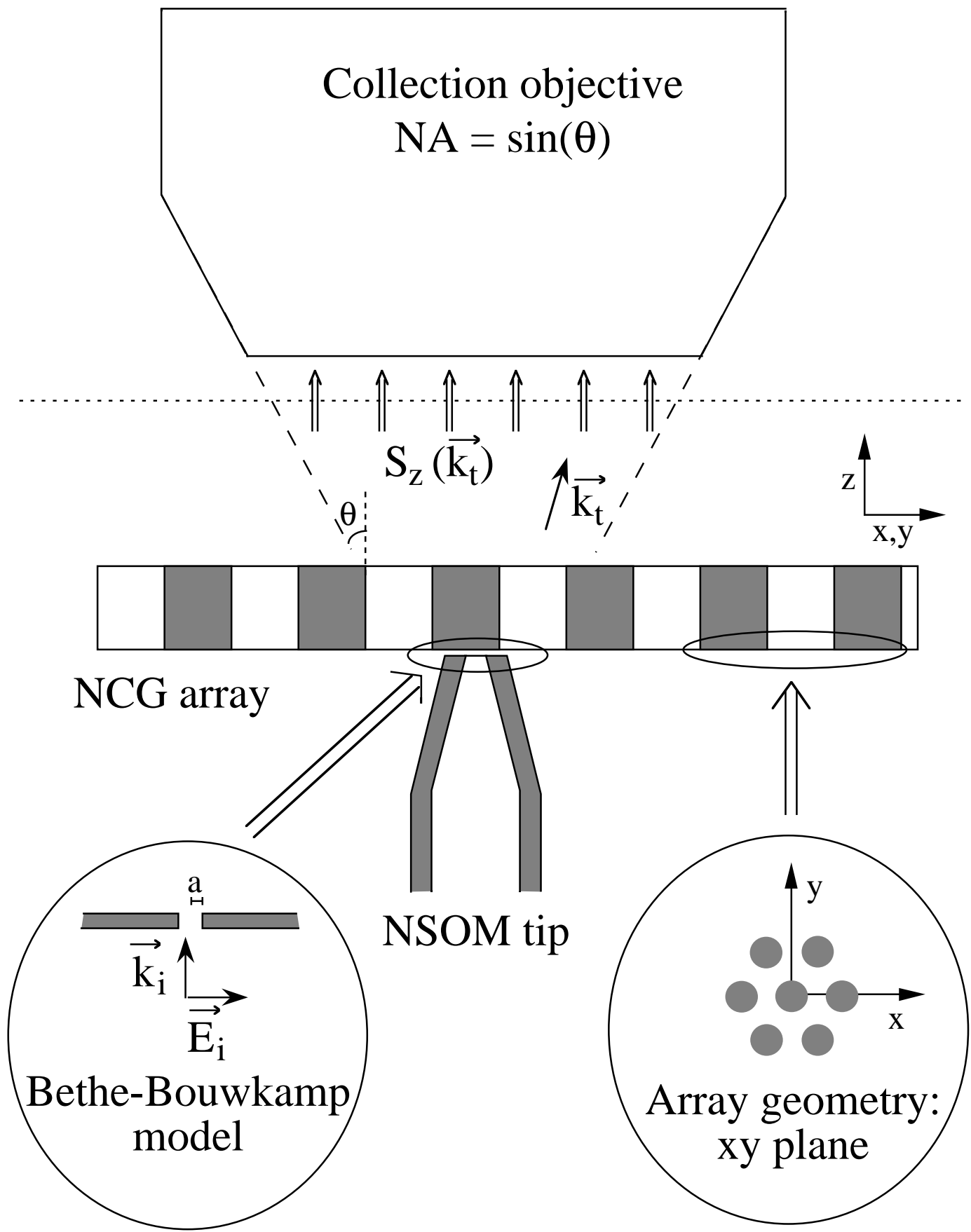
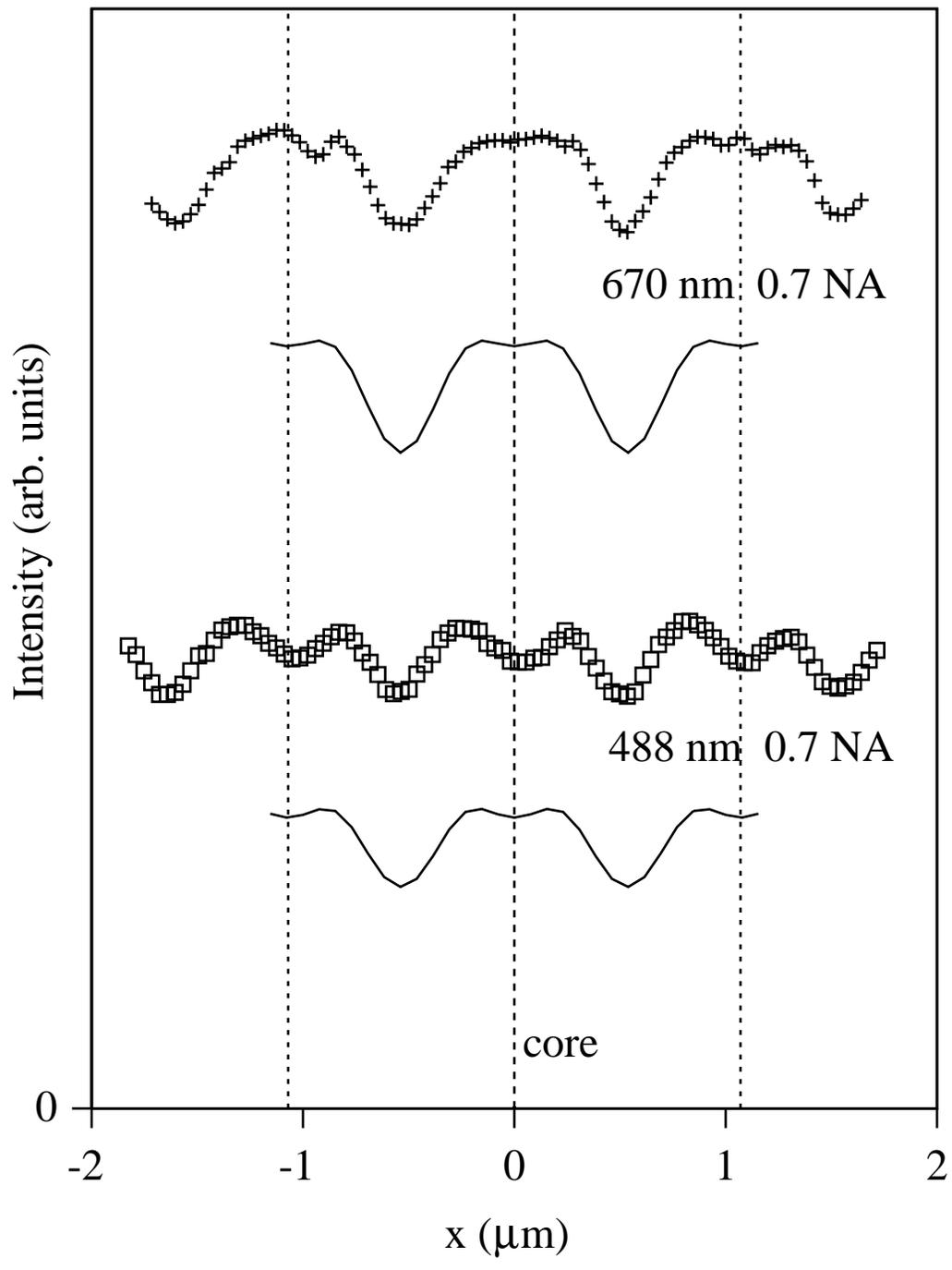
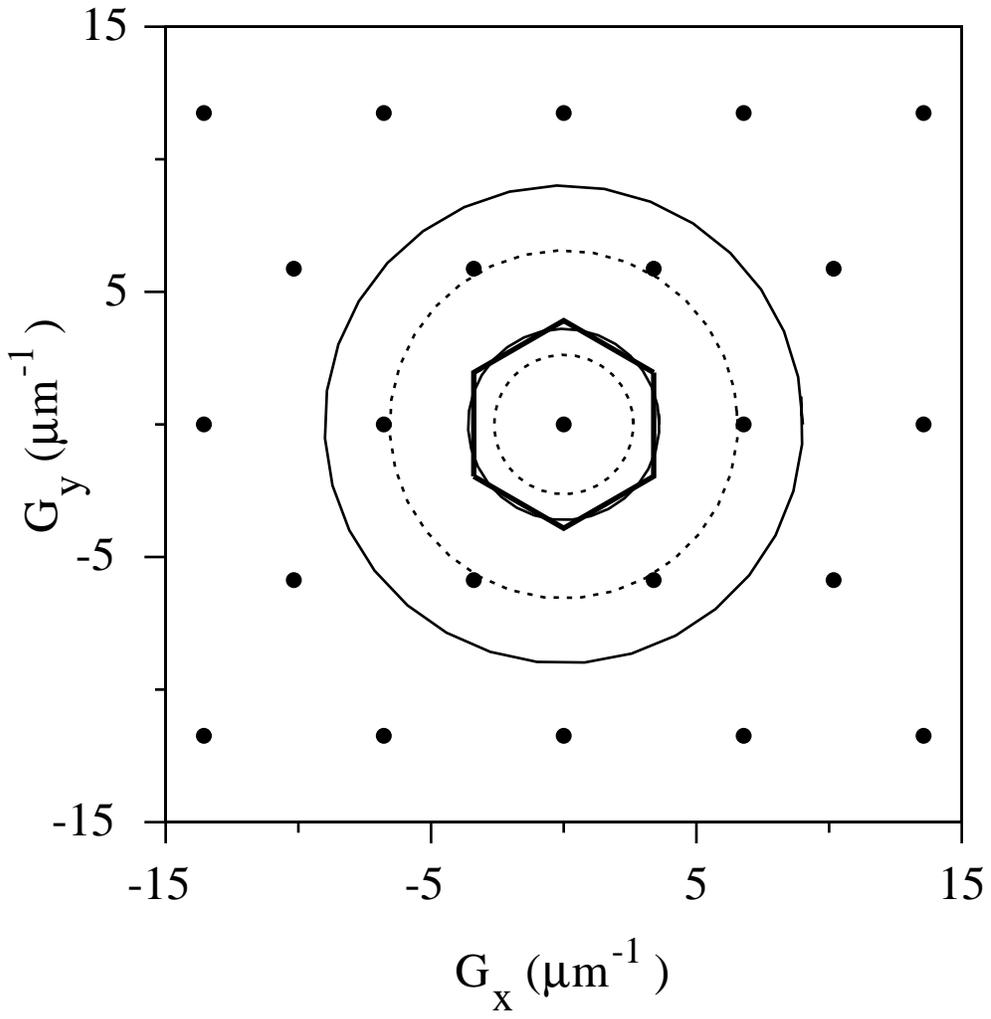
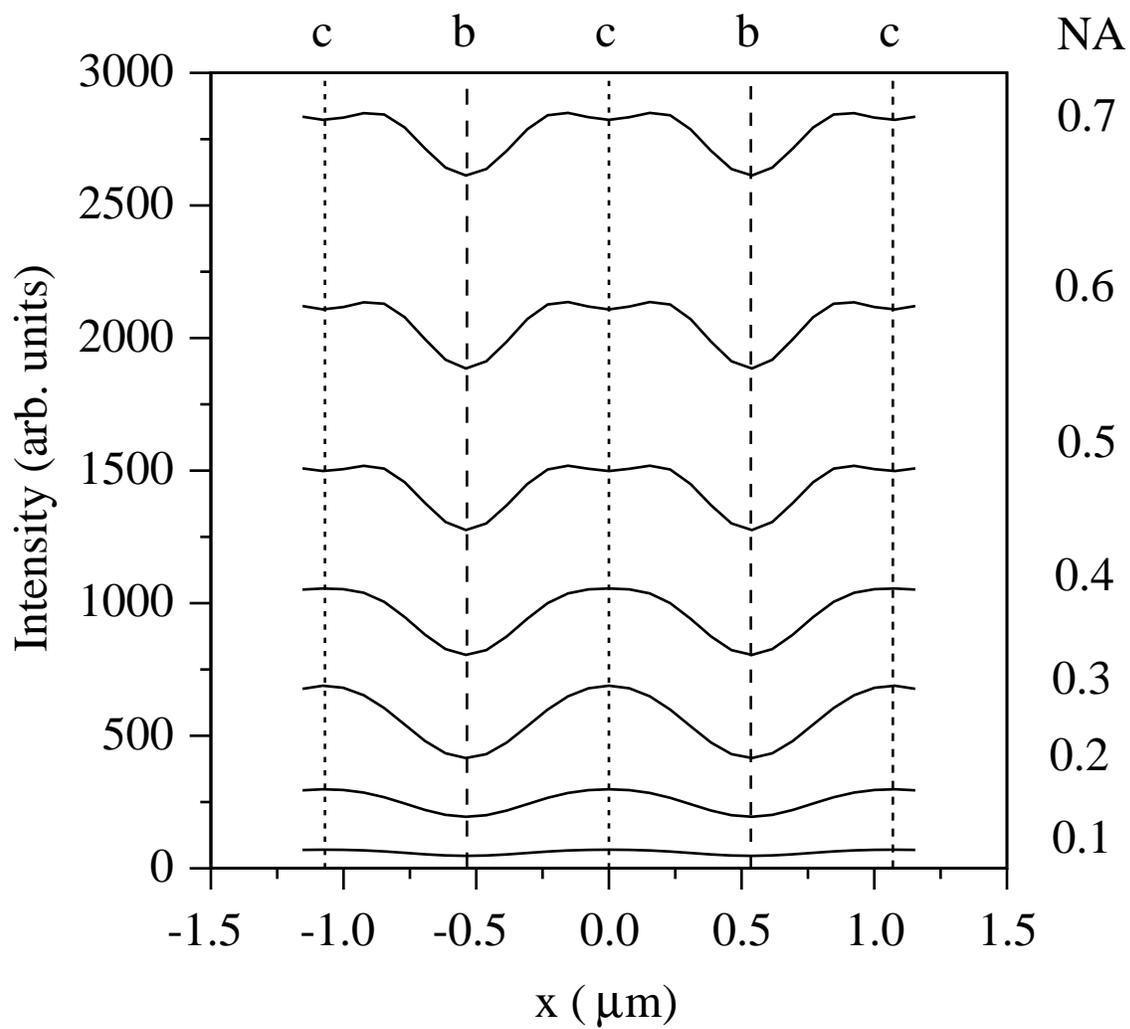
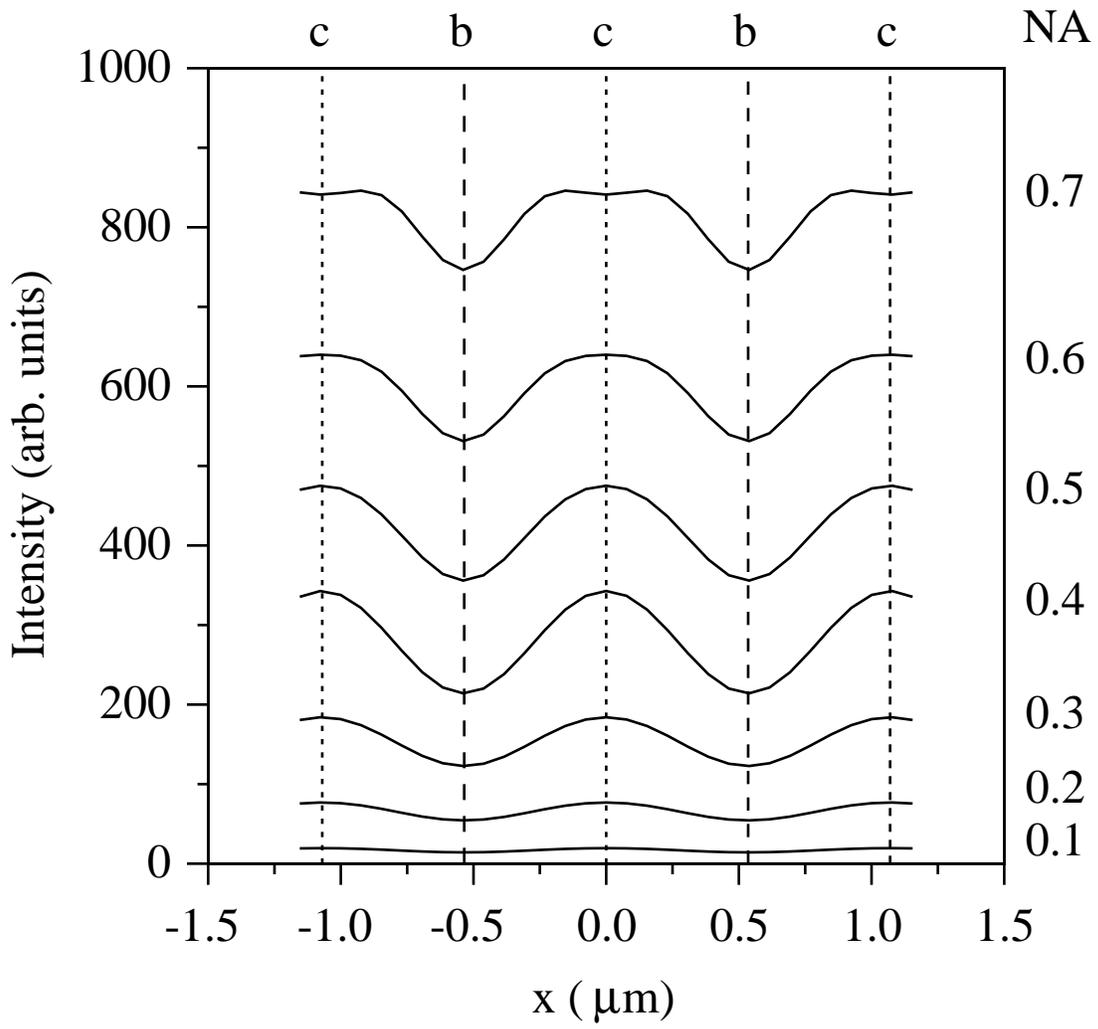
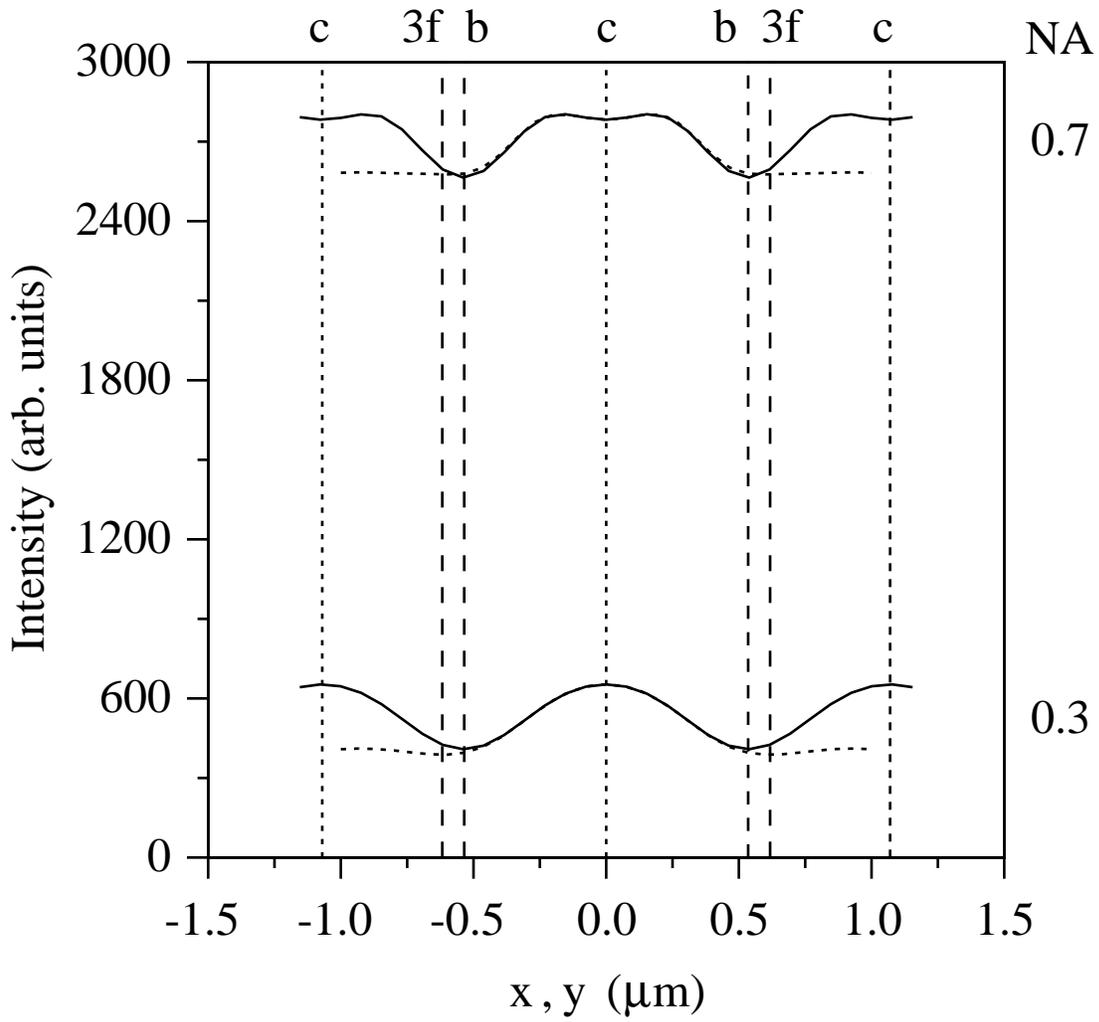
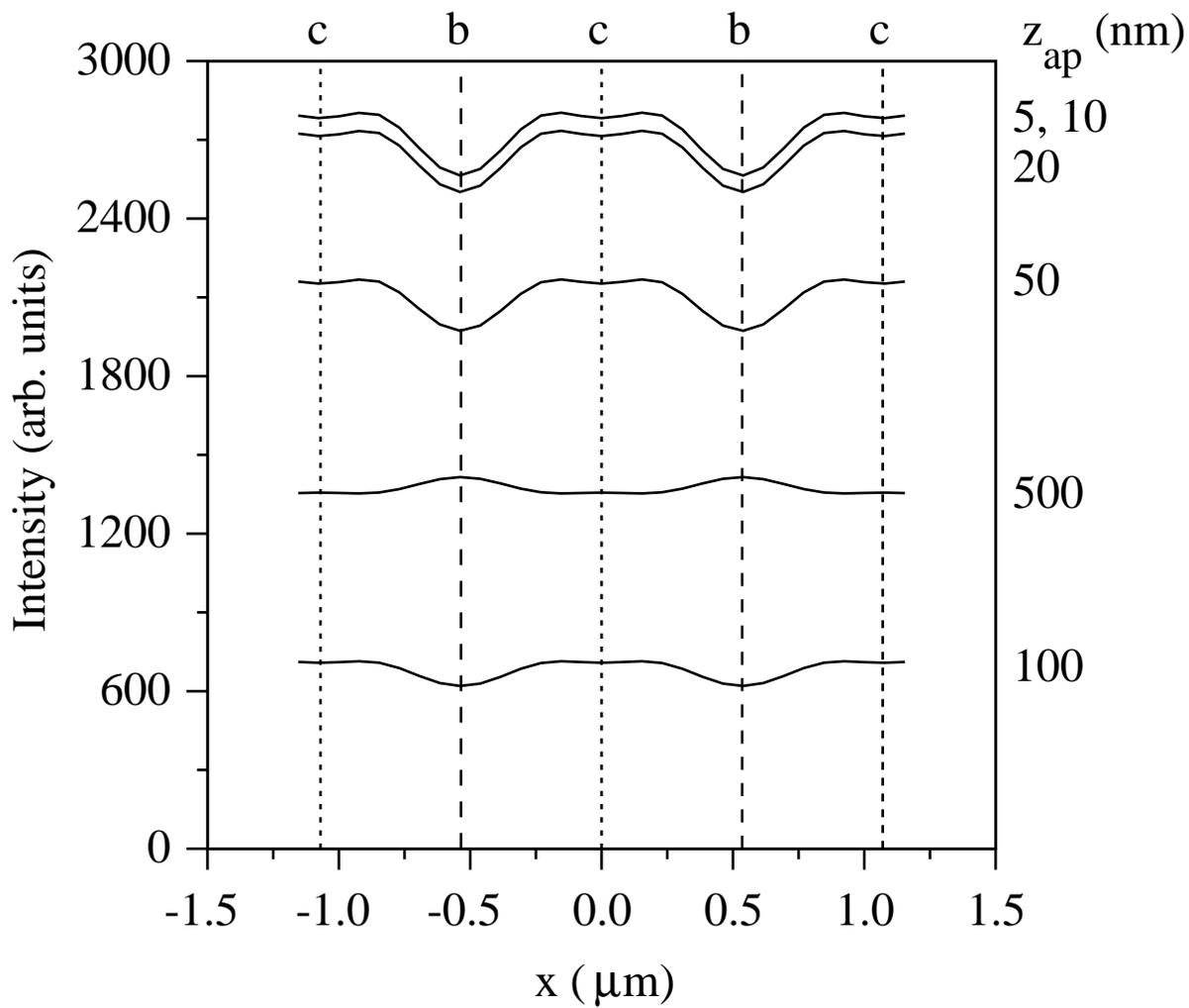
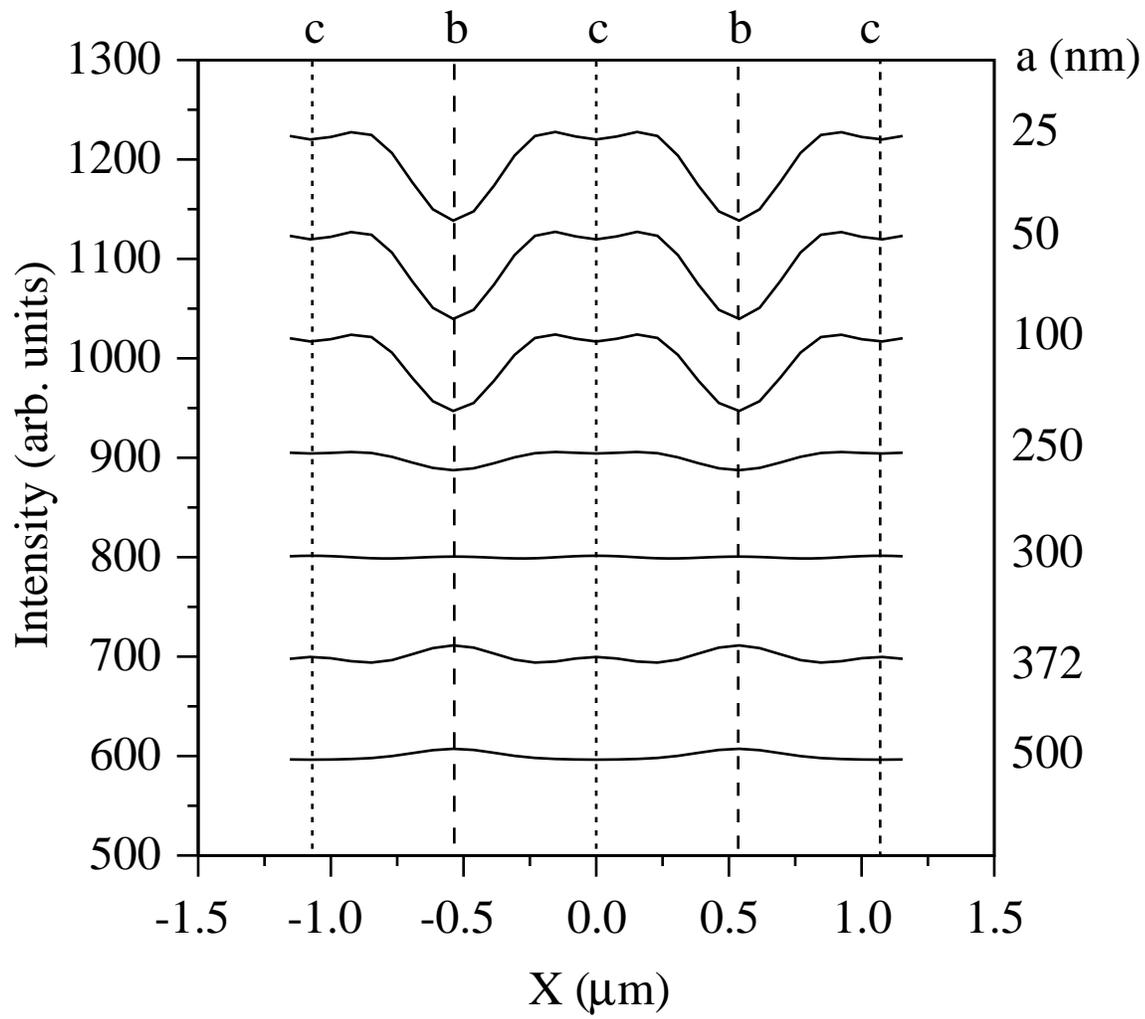
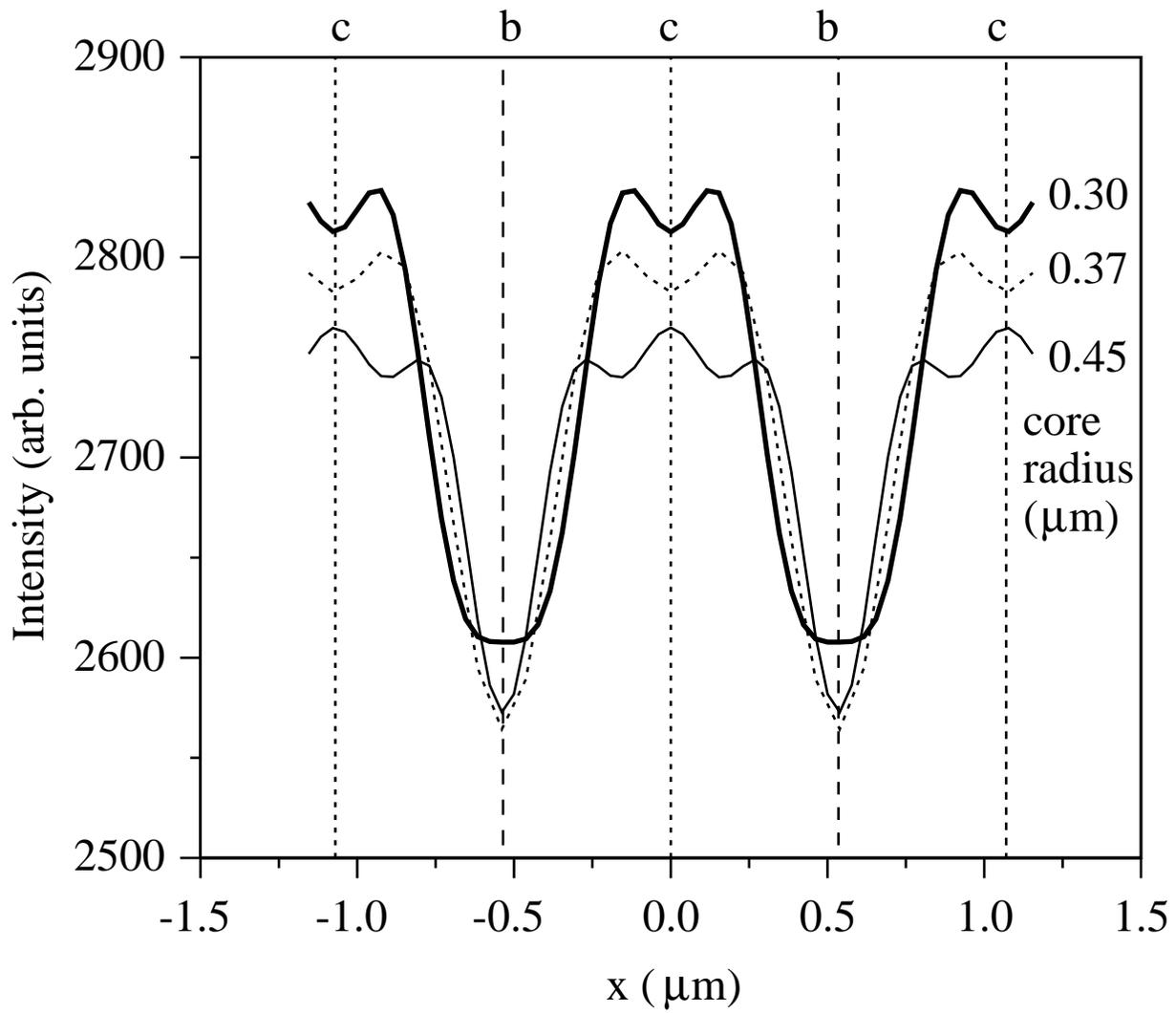
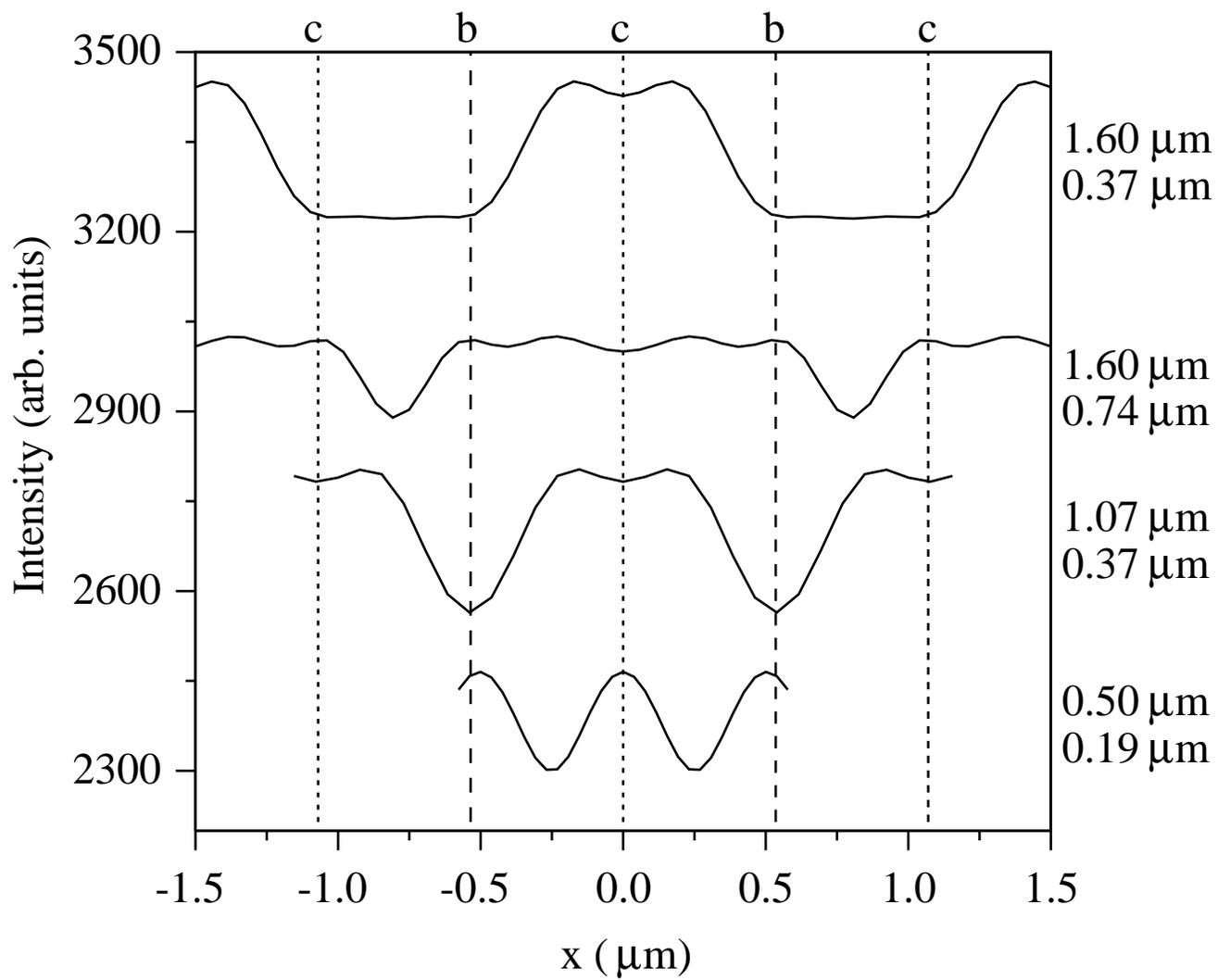
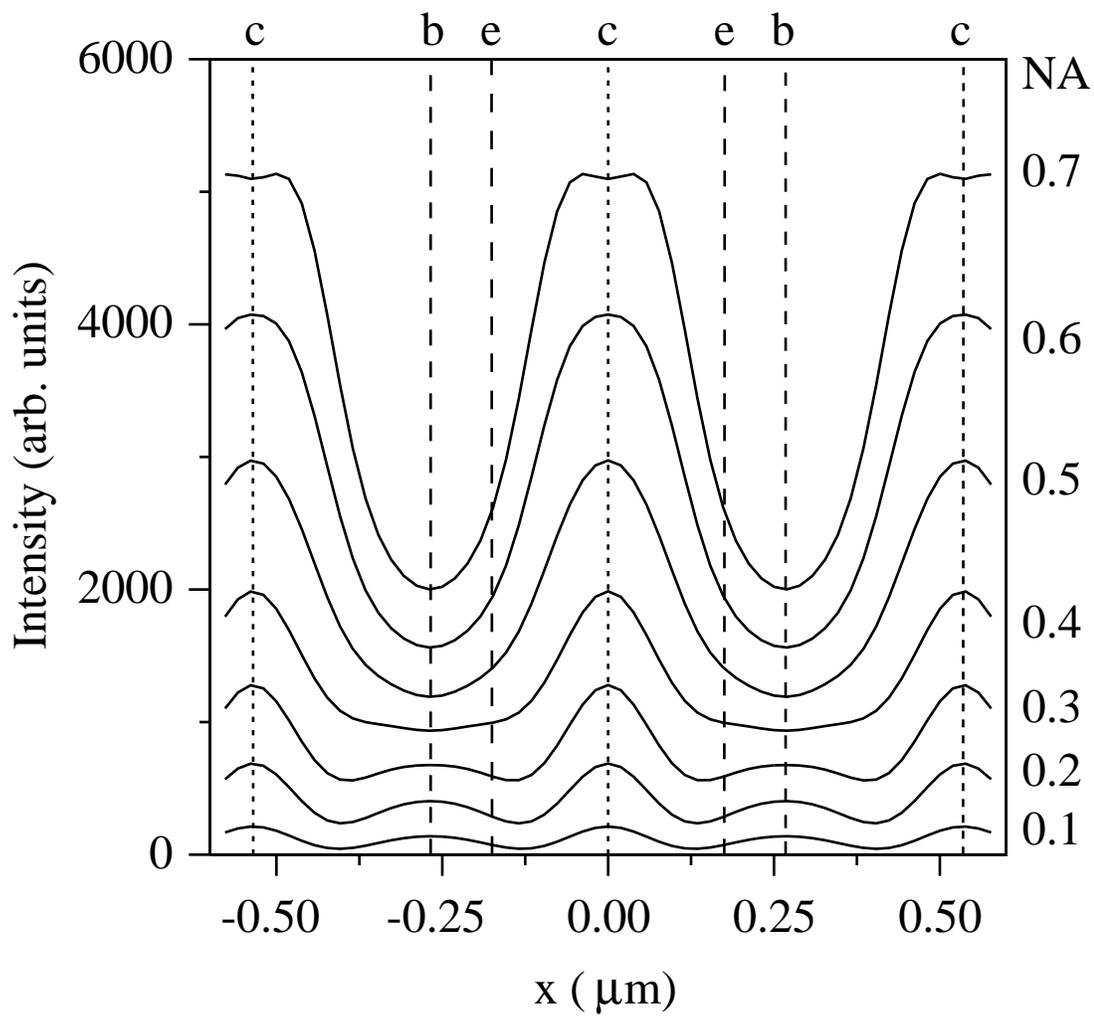
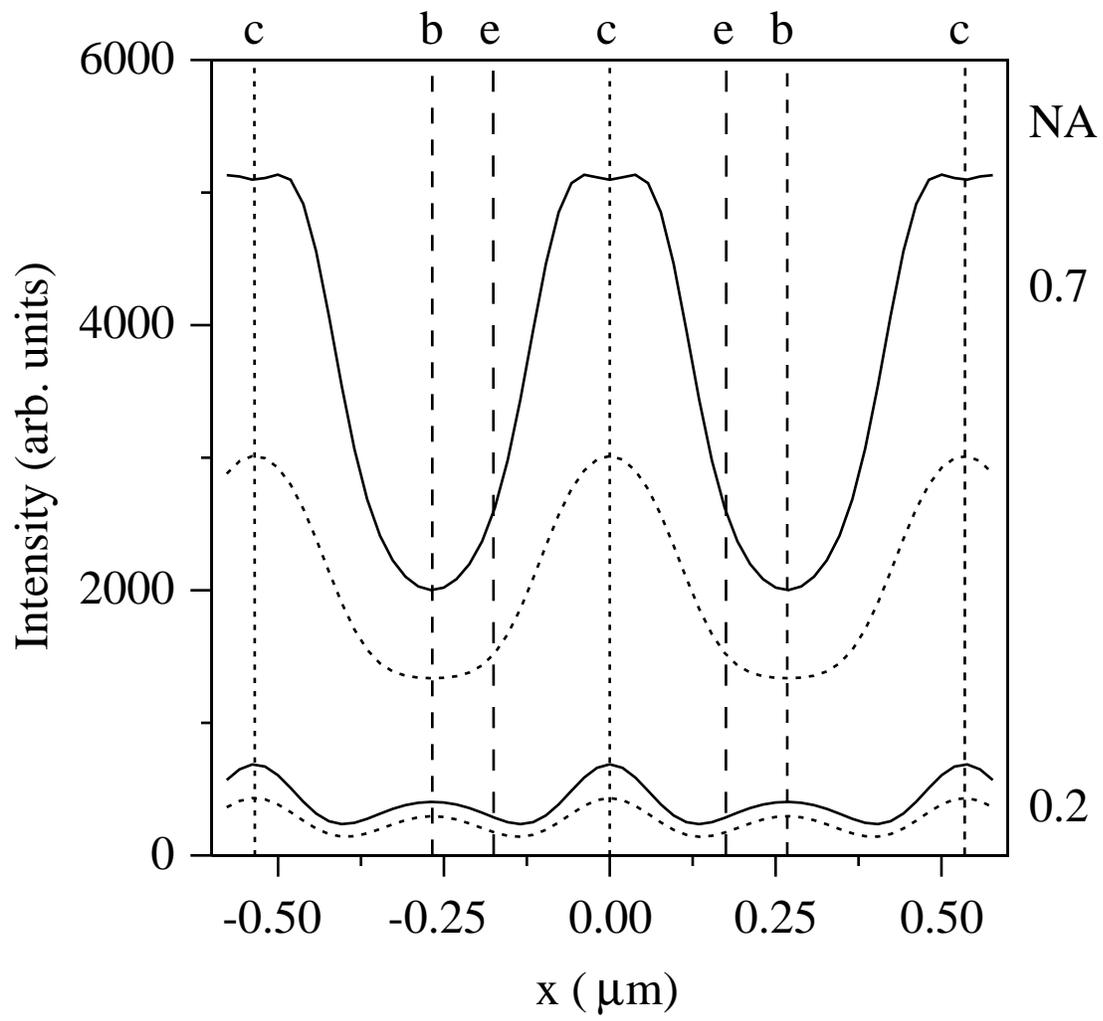
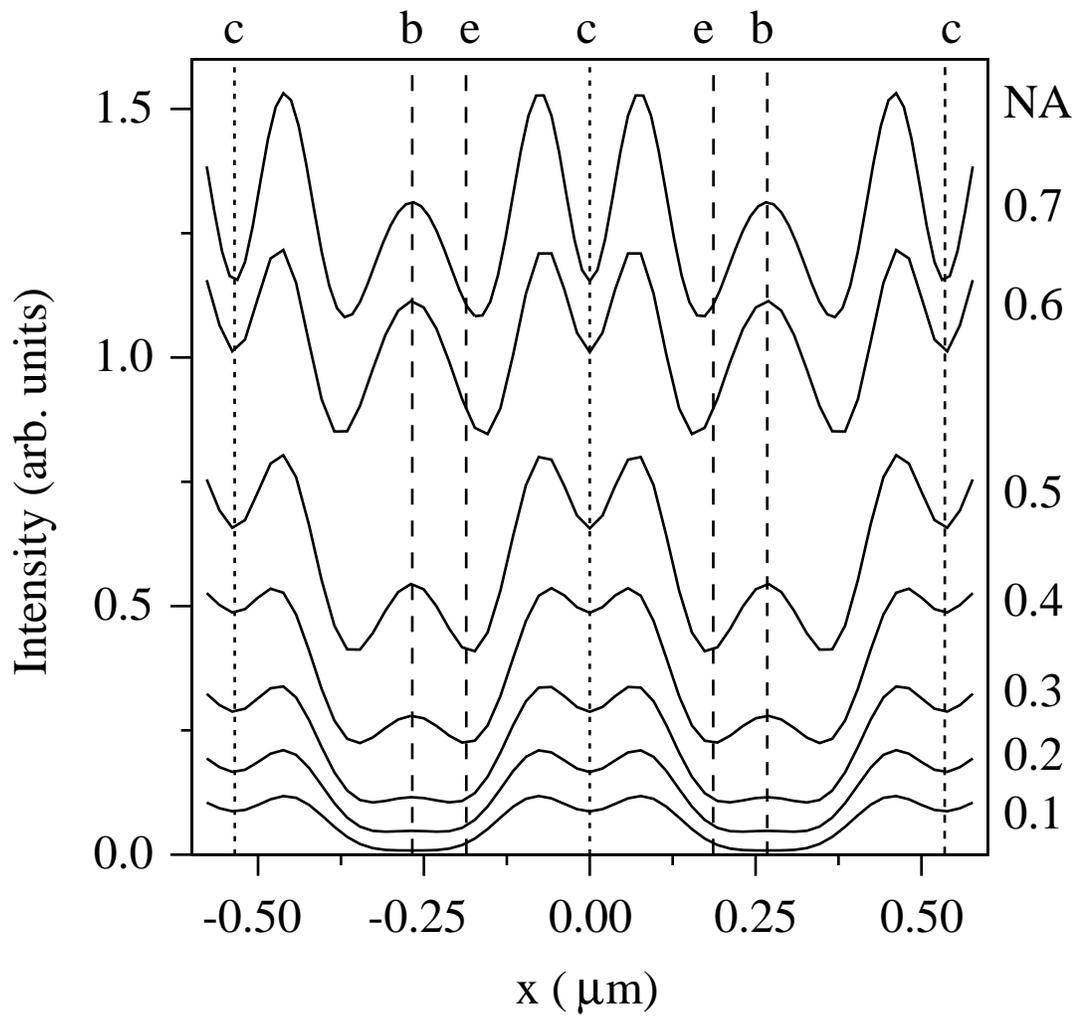
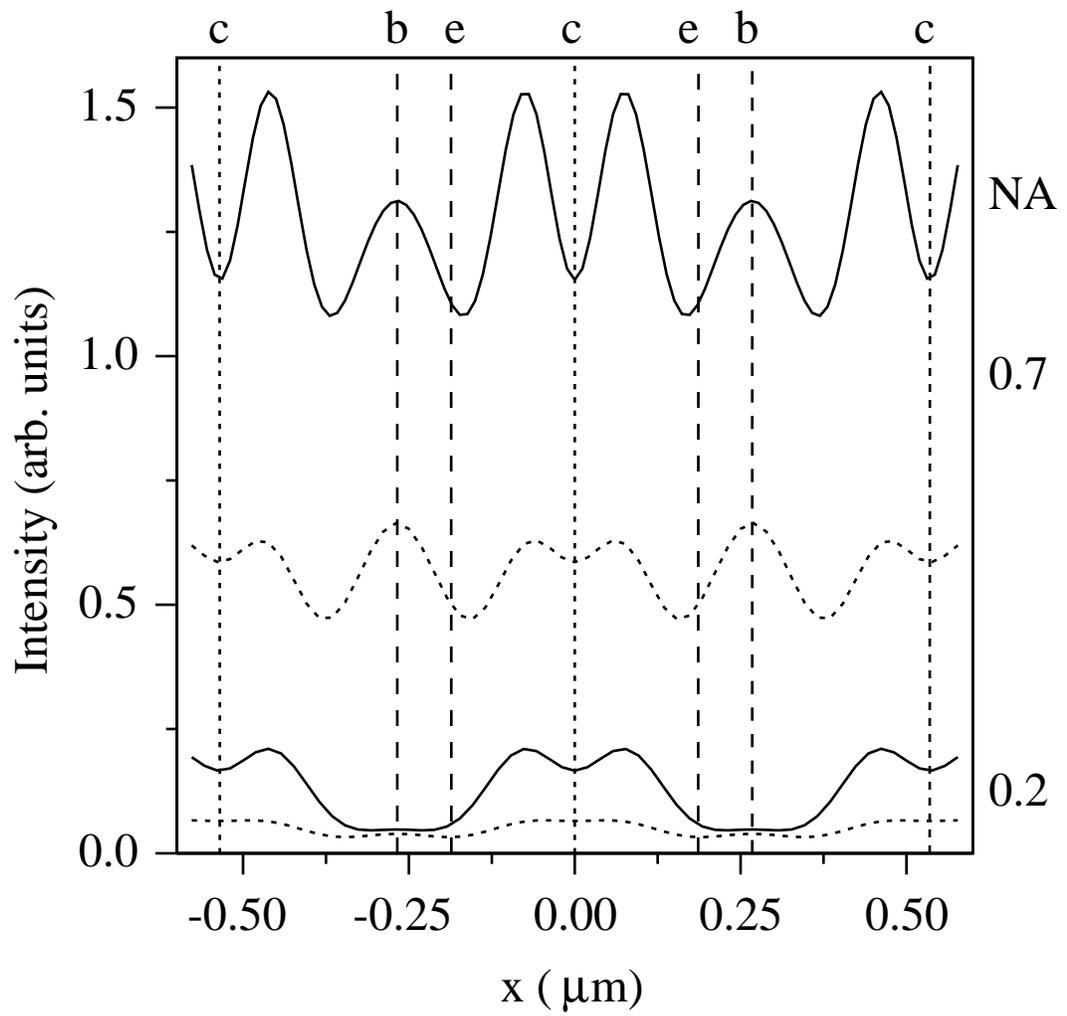
\end{document}